%% file: bare_adv.tex
\documentclass[10pt,journal,compsoc]{IEEEtran}

\usepackage{amsmath}
\usepackage{algorithm}
\usepackage[noend]{algpseudocode}

\usepackage{textcomp}

\usepackage{amsthm}
\usepackage{mathtools}
\usepackage{dsfont}

\usepackage{zref-savepos}

\usepackage{subcaption}
\usepackage{setspace}
\usepackage{footnote}
\usepackage[utf8]{inputenc} 
\usepackage[T1]{fontenc}    
\usepackage{hyperref}       
\usepackage{url}            
\usepackage{booktabs}       
\usepackage{amsfonts}       
\usepackage{nicefrac}       
\usepackage{microtype}      
\usepackage{xcolor}         
\usepackage{svg}
\usepackage{comment}
\usepackage{pdfpages}

\usepackage{graphicx,wrapfig,lipsum}
\usepackage[labelfont=bf]{caption} 
\usepackage{multirow}
\usepackage{color,soul}
\usepackage{makecell}
\usepackage{enumitem}
\usepackage{float}
\ifCLASSOPTIONcompsoc
  \usepackage[nocompress]{cite}
\else
  \usepackage{cite}
\fi
\ifCLASSINFOpdf
\else
\fi

\newcommand{\sysname}[0]{\textsc{Swift}}

\begin{document}
%
\title{\sysname{}: Expedited Failure Recovery for Large-scale DNN Training}
%
%
%
%

\author{Yuchen~Zhong, 
        Guangming~Sheng, 
        Juncheng~Liu, 
        Jinhui~Yuan, 
        and Chuan~Wu,~\IEEEmembership{Senior Member,~IEEE}
\IEEEcompsocitemizethanks{\IEEEcompsocthanksitem Yuchen Zhong, Guangming Sheng, and Chuan Wu are with the University of Hong Kong, Hong Kong, China. \protect\\ 
Email: \{yczhong, gmsheng, cwu\}@cs.hku.hk.
\IEEEcompsocthanksitem Juncheng Liu and Jinhui Yuan are with OneFlow Inc., Beijing, China. \protect\\
Email: \{liujuncheng, yuanjinhui\}@oneflow.org.
}
}

%
%

\markboth{}%
{Zhong \MakeLowercase{\textit{et al.}}: \sysname{}: Expedited Failure Recovery for Large-scale DNN Training}
%



\IEEEtitleabstractindextext{%
\begin{abstract}
As the size of deep learning models gets larger and larger, training takes longer time and more resources, making fault tolerance more and more critical. Existing state-of-the-art methods like CheckFreq and Elastic Horovod need to back up a copy of the model state (i.e., parameters and optimizer states) in memory, which is costly for large models and leads to non-trivial overhead. This paper presents \sysname{}, a novel recovery design for distributed deep neural network training that significantly reduces the failure recovery overhead without affecting training throughput and model accuracy. Instead of making an additional copy of the model state,  \sysname{} resolves the inconsistencies of the model state caused by the failure and exploits the replicas of the model state in data parallelism for failure recovery. We propose a logging-based approach when replicas are unavailable, which records intermediate data and replays the computation to recover the lost state upon a failure. The re-computation is distributed across multiple machines to accelerate failure recovery further. We also log intermediate data selectively, exploring the trade-off between recovery time and intermediate data storage overhead. Evaluations show that \sysname{} significantly reduces the failure recovery time and achieves similar or better training throughput during failure-free execution compared to state-of-the-art methods without degrading final model accuracy.  \sysname{} can also achieve up to 1.16x speedup in total training time compared to state-of-the-art methods.
\end{abstract}

}

\maketitle

\IEEEdisplaynontitleabstractindextext

%
\IEEEpeerreviewmaketitle


\input{introduction.tex}

\input{background_and_motivation.tex}

\input{overview.tex}

\input{system_design.tex}
\input{implementation.tex}
\input{evaluation.tex}

\input{related_work.tex}

\section{Conclusion}
This paper presents \sysname{}, a novel design that expedites failure recovery in distributed DNN training. \sysname{} exploits redundancies in data-parallel training for failure recovery and resolves the crash-consistency problem with update-undo. \sysname{} advocates logging for pipeline-parallel training, which records inter-machine intermediate data at runtime and limits the computation graph to be re-executed to those on the failed workers. We also design parallel recovery to expedite recovery further and explore the trade-off between recovery time and space overhead with selective logging. Compared to state-of-the-art approaches, extensive evaluations show that \sysname{} significantly accelerates failure recovery without affecting training throughput and model accuracy.

\ifCLASSOPTIONcaptionsoff
  \newpage
\fi



%



\bibliographystyle{IEEEtran}
\bibliography{citation}

\input{appendix}

\end{document}

%% file: introduction.tex

\ifCLASSOPTIONcompsoc
\IEEEraisesectionheading{\section{Introduction}\label{sec:introduction}}
\else
\section{Introduction}
\label{sec:introduction}
\fi

\IEEEPARstart{L}{arger} and larger deep neural networks (DNNs) have recently emerged for improved model performance~\cite{devlin2018bert, shoeybi2019megatron, brown2020language}. Large DNN model training jobs typically use many accelerators (e.g., GPUs) and have long-running times~\cite{jeon2019analysis}. For example,  training a GPT-3 model~\cite{brown2020language} on 1024 A100 GPUs is estimated to take more than one month~\cite{narayanan2021efficient}. 

Job failures are common in a GPU training cluster~\cite{jeon2019analysis}. For example, machine crashes and network failures happen occasionally, or higher priority jobs take up resources~\cite{jeon2019analysis}. In these cases, the distributed DNN training job is interrupted, resulting in loss of the DNN model state (i.e., model parameters and optimizer states) and failure of the training job. Failures are more severe for large DNN model training jobs: increasing the number of machines will inevitably lead to an increased chance of failure; training large models takes days to months, making it more likely for failures to happen during the course. Recent works also echo this~\cite{maeng2021understanding, eisenman2022check}.

Global checkpointing is the \emph{de facto} method for fault tolerance in deep learning (DL) frameworks~\cite{abadi2016tensorflow, paszke2019pytorch}. The training job periodically checkpoints the entire model state. All workers restart from the latest checkpoint when the job fails. Depending on the checkpointing frequency, this often results in several hours of lost computation time~\cite{mohan2021checkfreq}. CheckFreq~\cite{mohan2021checkfreq} achieves more frequent checkpoints by splitting the operation into two phases: first, the model state is copied in the GPU memory, called a \textit{snapshot}, or to the CPU memory if the GPU memory is insufficient; in the second phase, the snapshot is written to the disk asynchronously. Elastic Horovod~\cite{elastichorovod}, a framework for elastic training, takes a similar approach, but without the second phase. The reason is that Elastic Horovod assumes distributed data-parallel training, where each worker maintains a replica of the model state; during failure recovery, one of the surviving workers broadcasts the snapshot to other workers, and all workers restart training from the snapshot. Taking a snapshot is necessary for Elastic Horovod to prevent a corrupted state: if a worker crashes during the parameters update, the other workers are in an awkward situation - some parameters are updated while the others are not. We identify this problem as the \textit{crash-consistency problem} (\S\ref{sec2.3:crash_consistency problem}). However, as we shall see in \S\ref{sec2.2:fault_tolerance_in_deep_learning}, for large DNN models, both methods can slow down the training due to the overhead of snapshotting. 

This paper studies a better failure resilience design for distributed DNN training that significantly reduces the recovery overhead without affecting training throughput and final model accuracy. One of our key observations is that many of the optimizers used for model state updates in DNN training are mathematically \textit{invertible}. For example, stochastic gradient descent (SGD) only involves linear operators such as element-wise addition and scalar multiplication, and the inverse operators are straightforward. In case of a crash-consistency problem in distributed data-parallel training, we can restore the model states of the surviving workers to a consistent state by \textit{undoing the update} of the updated parameters (\S\ref{sec:undo_update}). Therefore, we do not need to snapshot periodically as CheckFreq and Elastic Horovod do, reducing the overhead during failure-free execution to zero (except for periodic checkpoints). Since this approach exploits replicas of the model state in surviving workers for recovery, we name this recovery method \textit{replication-based recovery}. 

However, replicas are not always available, even with data parallelism. For instance, some prior works advocate data parallelism only across multiple GPUs on the same machine to leverage high-speed intra-server interconnects such as NVLink~\cite{fan2021dapple, narayanan2021efficient} to accelerate gradient synchronization. All replicas would be lost in the event of a machine failure. 

\begin{figure}[t]
    \centering
    \begin{subfigure}[b]{\columnwidth}
        \centering
        \includegraphics[width=\columnwidth]{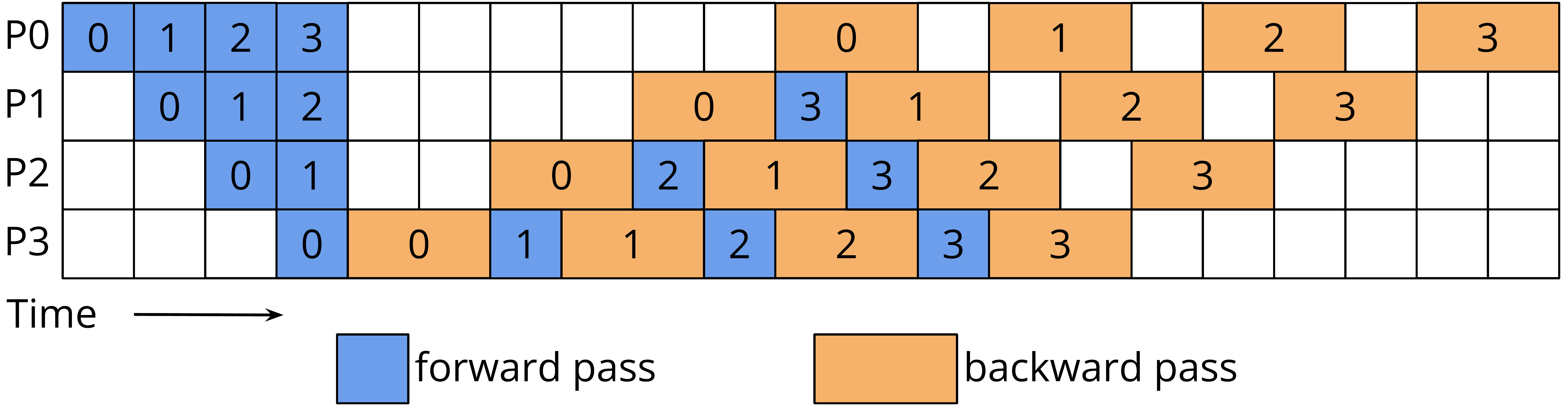}
        \caption{Pipeline parallelism with four devices (P0, ..., P3) with 1F1B schedule. The number in a block indicates the micro-batch being processed. The white blocks represent bubble time.}
        \label{fig:pipeline_parallelism_1f1b}
    \end{subfigure}
    \hfill
    \begin{subfigure}[b]{\columnwidth}
        \centering
        \includegraphics[width=\columnwidth]{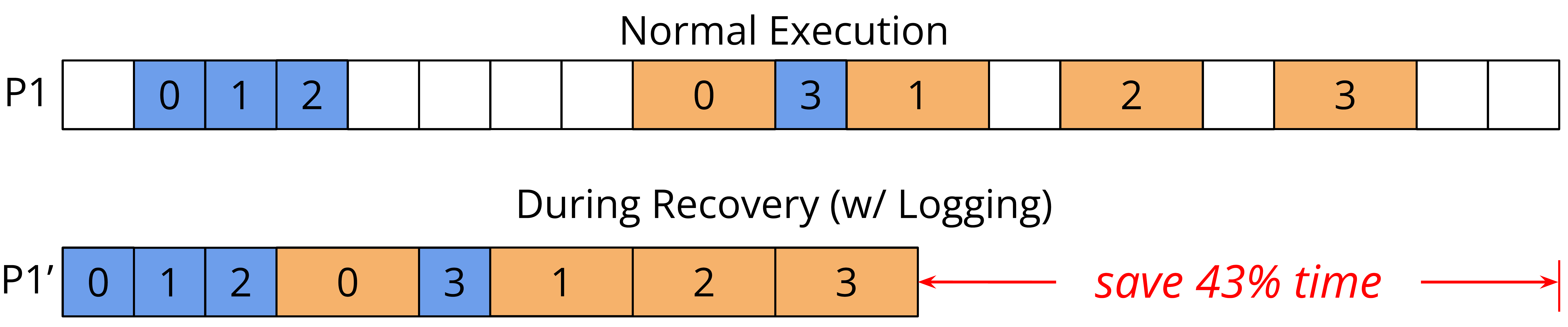}
        \caption{An example showing that logging can expedite recovery after failure, where worker P1 in (a) has failed and is replaced by a new worker P1'. During normal execution, P1 needs to wait for intermediate activations or gradients from other workers. When recovering the lost state of P1, P1' can read the logged data (activations and gradients from other workers) and directly do the re-computation (without the bubble time as in normal execution), achieving much reduced recovery time than checkpoint recovery.}
    \label{fig:logging_pipeline_comparison}
  \end{subfigure}
  \caption{Pipeline parallelism and logging example.}
\end{figure}

We then investigate another fundamental approach for fault tolerance in distributed systems - logging, which has been widely explored in data processing systems~\cite{zaharia2012resilient, wang2019lineage, hwang2005high, shen2014fast}. We introduce \textit{logging-based recovery} (\S\ref{sec:logging}) for pipeline-parallel training. In pipeline parallelism, workers form a chain topology and pass intermediate activations or gradients to the successor or predecessor worker using point-to-point communication. Figure~\ref{fig:pipeline_parallelism_1f1b} illustrates One-Forward-One-Backward (1F1B) pipeline schedule~\cite{narayanan2019pipedream} (\S\ref{sec2.1:distributed_dnn_training}). With logging, each worker locally records all outgoing data to the adjacent workers on the other machine. Upon a failure, the replacements of the failed workers retrieve the logging data and replay the computation to recover the lost state. Moreover, we spread the logging data to surviving workers to have them assist in recovery (\S\ref{sec:4.2:parallel_recovery}). Logging \textit{limits the recovery scope from the complete computation graph across all workers to the computation graph on the failed workers}, thus reducing the recovery time compared to global checkpointing. An example is given in Figure~\ref{fig:logging_pipeline_comparison}. To the best of our knowledge, we are the first to bring logging into distributed DL systems for failure resilience.

However, logging-based recovery brings unique challenges. Logging needs to be done constantly during DNN training, and the overhead in runtime and space can be prohibitive. Once a piece of logged data is missing, the original state cannot be recovered precisely. To reduce the runtime overhead of logging, we only log inter-machine communication data since failures often occur at machines rather than at individual workers on machines. Moreover, we perform logging asynchronously, storing the data in the background. We further utilize workers’ idle time (i.e., bubble time in pipeline parallelism) to do logging. This way, logging is \textit{off the critical path} (\S\ref{sec:4.1:basic_mechanism}). To control the space overhead at a manageable level, we devise an algorithm to select only a subset of machines to log intermediate data (\S\ref{sec:4.3:selective_logging}). Selective logging trades the recovery time for space consumption.

We design and implement \sysname{}, including replication-based and logging-based recovery for expedited failure recovery. Our key contributions are summarized below: 
    
$\triangleright$ We propose a novel mechanism called update-undo that resolves model state inconsistencies caused by the failure and enables failure recovery using replicas of the model state in data parallelism without creating additional copies.

$\triangleright$ We propose to use the logging method to achieve expedited failure recovery in pipeline parallelism. We use asynchronous logging, logging during the bubble time, and selective logging to reduce the runtime and space overhead.

$\triangleright$ We implement \sysname{} in PyTorch and demonstrate its benefits on distributed training of large DNN models. For replication-based recovery in training Wide-ResNet-50~\cite{zagoruyko2016wide}, \sysname{} reduces recovery time by 98.9\%, 98.1\%, and 98.1\% compared to global checkpointing, CheckFreq and Elastic Horovod, respectively. For logging-based recovery in training BERT~\cite{devlin2018bert} and ViT~\cite{dosovitskiy2020image}, \sysname{} reduces recovery time by 57.3\% and 76.3\% compared to global checkpointing, respectively. Using traces collected in our experiments, we show that \sysname{} can achieve up to 1.16x speedup in total training time compared to state-of-the-art methods. We have open-sourced \sysname{} at \url{https://github.com/jasperzhong/swift}. 

%% file: background_and_motivation.tex
\section{Background and Motivation}
\subsection{Distributed DNN Training}
\label{sec2.1:distributed_dnn_training}
We focus on \textit{synchronous} distributed DNN training, where many workers on multiple machines collectively work on the latest DNN model iteratively. Each training iteration contains a forward computation pass (to compute a loss) and a backward pass (to compute the gradients), and the gradients computed are used for the model update~\cite{chen2016revisiting}. Synchronous training ensures better model accuracy than asynchronous training and is thus popular for large-scale DNN training~\cite{chen2016revisiting, goyal2017accurate, you2019large, narayanan2021efficient}. 

\vspace{1mm}
\noindent\textbf{Data parallelism}
is the most widely used paradigm for distributed DNN training~\cite{goyal2017accurate, you2019large}. Input data is partitioned across workers. Each worker has a model replica and computes local gradients on a subset of data. Gradient synchronization is performed among workers in each iteration to ensure the consistency of model replicas. 

\vspace{1mm}
\noindent\textbf{Operator parallelism} is a solution to handle large DNNs by splitting an operator in a DNN model among multiple workers along non-batch axes~\cite{jia2019beyond}. Communication is needed to fetch the input data from other workers~\cite{shoeybi2019megatron}.

\vspace{1mm}
\noindent\textbf{Pipeline parallelism} splits a mini-batch into smaller micro-batches and pipelines them to the DNN model stages hosted on different workers so that workers can process different micro-batches simultaneously~\cite{huang2019gpipe, narayanan2019pipedream, narayanan2021memory, fan2021dapple}. Point-to-point communication is performed between workers hosting neighbor stages to transfer intermediate activations. Synchronous pipeline parallelism schedules like GPipe~\cite{huang2019gpipe} and One-Forward-One-Backward (1F1B)~\cite{narayanan2021memory} flush the pipeline in each iteration, i.e., worker waiting for all in-flight micro-batches of the iteration to complete before moving on to the next iteration. Despite better model accuracy, pipeline flush causes worker idling (i.e., \textit{bubbles}) in pipeline execution~\cite{huang2019gpipe, narayanan2021memory, fan2021dapple, narayanan2021efficient}. For GPipe and 1F1B, the ratio of the bubble time is $(p-1)/(m+p-1)$, where $p$ is the number of stages and $m$ is the number of micro-batches~\cite{narayanan2021efficient}. For the example shown in Figure~\ref{fig:pipeline_parallelism_1f1b}, the ratio of the bubble time is $3/7$. This paper adopts 1F1B~\cite{narayanan2021memory} because it has the same bubble time ratio but lower peak memory usage than GPipe~\cite{huang2019gpipe}. Note that our approach is not limited to 1F1B.

\begin{figure}[t]
    \centering
    \includegraphics[width=\columnwidth]{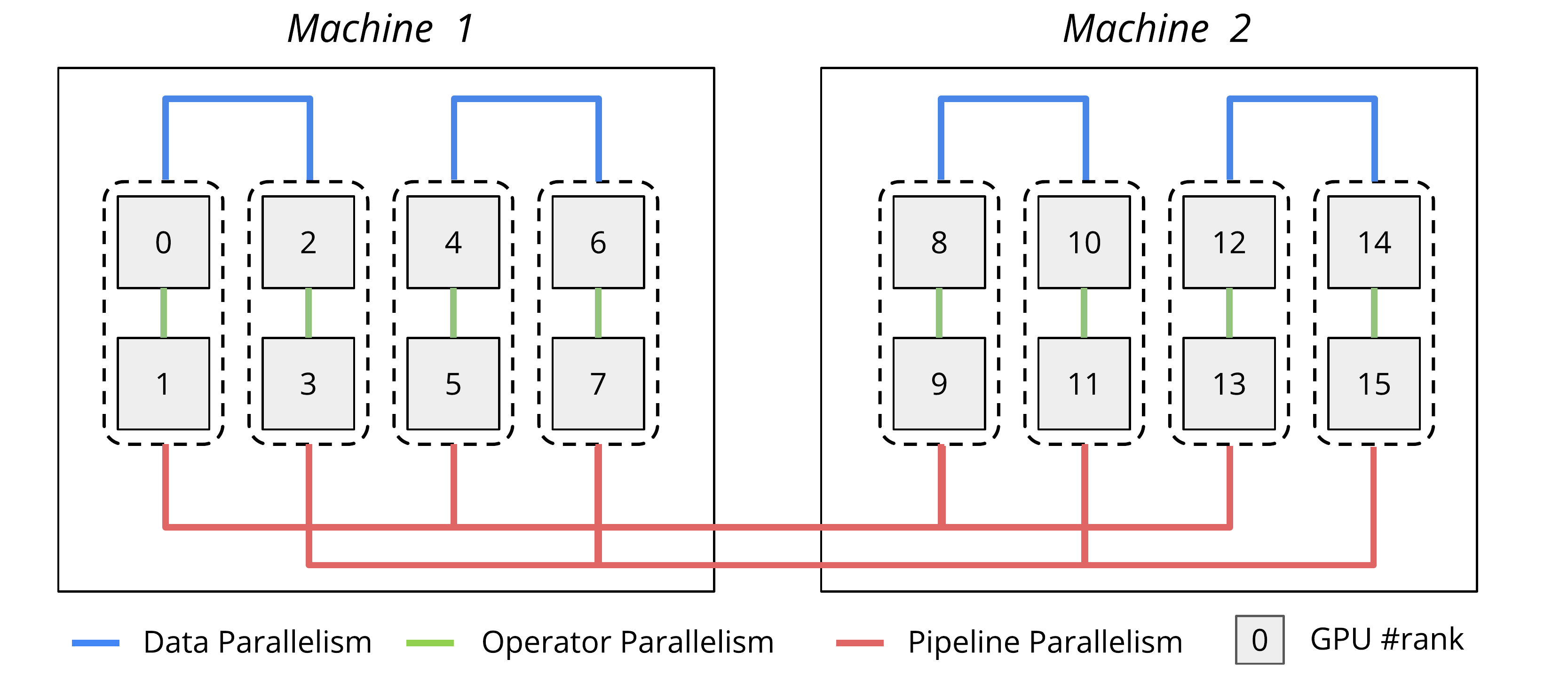}
    \caption{A hand-optimized 3D parallelism plan in Megatron-LM, using 16 GPUs on two machines. The DNN model is split into four pipeline-parallel stages, each stage is partitioned onto two GPUs for operator parallelism and each stage has a replica. Replicas of a stage are on the same machine.}
    \label{fig:3d_parallelism}
\end{figure}

Recent works combine the three parallelism paradigms, called \textit{3D parallelism}~\cite{shoeybi2019megatron, fan2021dapple, narayanan2021efficient, yuan2021oneflow, zheng2022alpa}. Figure~\ref{fig:3d_parallelism} shows a hand-optimized parallelism plan in Megatron-LM~\cite{shoeybi2019megatron, narayanan2021efficient}, a state-of-the-art training system for transformer language models. Although data parallelism is used in this example, the replicas reside on the same machine. If one machine fails, we lose the model state on that failed machine.

\subsection{Problems on Snapshotting Large Models}
\label{sec2.2:fault_tolerance_in_deep_learning}


\begin{figure}[t]
    \centering
    \includegraphics[width=\columnwidth]{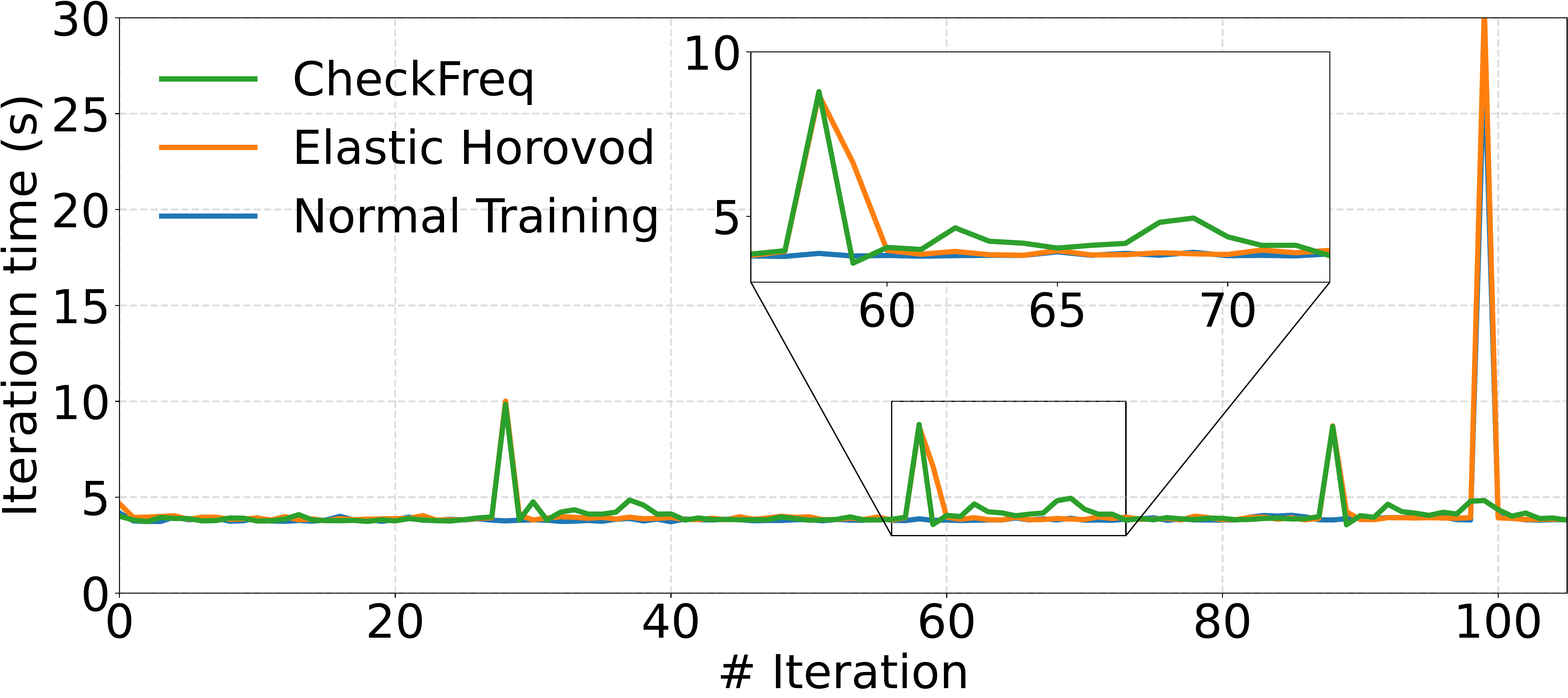}
    \caption{Training throughput of Wide-ResNet-50 during failure-free execution.}
    \label{fig:cdf_wide_resnet}
\end{figure}

CheckFreq~\cite{mohan2021checkfreq} and Elastic Horovod~\cite{elastichorovod}, state-of-the-art methods for fault tolerance, rely on the \textit{snapshot} operation. After updating the model state of the iteration, a copy of the model state (called a snapshot) is captured in GPU memory or copied to CPU memory if the GPU cannot hold it. Snapshotting can overlap with the next iteration's forward and backward pass. The next iteration of the update operation does not start until the snapshot operation is completed, leading to a checkpoint stall. However, DNN models have proliferated from millions to billions of parameters in recent years and become too large to fit into a single GPU~\cite{zero-infinity}. It became increasingly difficult to fit a complete snapshot on a single GPU. In that case, the snapshot is copied to the CPU. 

We experimentally find that snapshotting to CPU memory is costly for large models and reduces training throughput. We train an enlarged Wide-ResNet-50~\cite{zagoruyko2016wide} model with a model state size of 9.8GB using data parallelism on two machines using 8 32GB V100 GPUs. The snapshot operation needs to copy the model state to the CPU. The model setting, the training setting, and the settings of CheckFreq and Elastic Horovod are described in \S\ref{sec:experimental_setup}. During training, the GPU memory consumption reaches 30.4 GB, which cannot accommodate a snapshot. Figure~\ref{fig:cdf_wide_resnet} shows that at the time of snapshots (iterations 30, 60, and 90), the iteration time is significantly longer with CheckFreq and Elastic Horovod. After the snapshots, CheckFreq's iteration time is longer, showing that writing the snapshot to the disk also affects normal training. But global checkpointing causes large overhead (iteration 100) since it is synchronous. 
Interestingly, the checkpoint stall is indeed negligibly with CheckFreq, for only 0.2 milliseconds. This experiment shows that the snapshot operation can still incur non-trivial runtime overhead and slow down the training process.

\begin{figure}[t]
     \centering
    \includegraphics[width=\columnwidth]{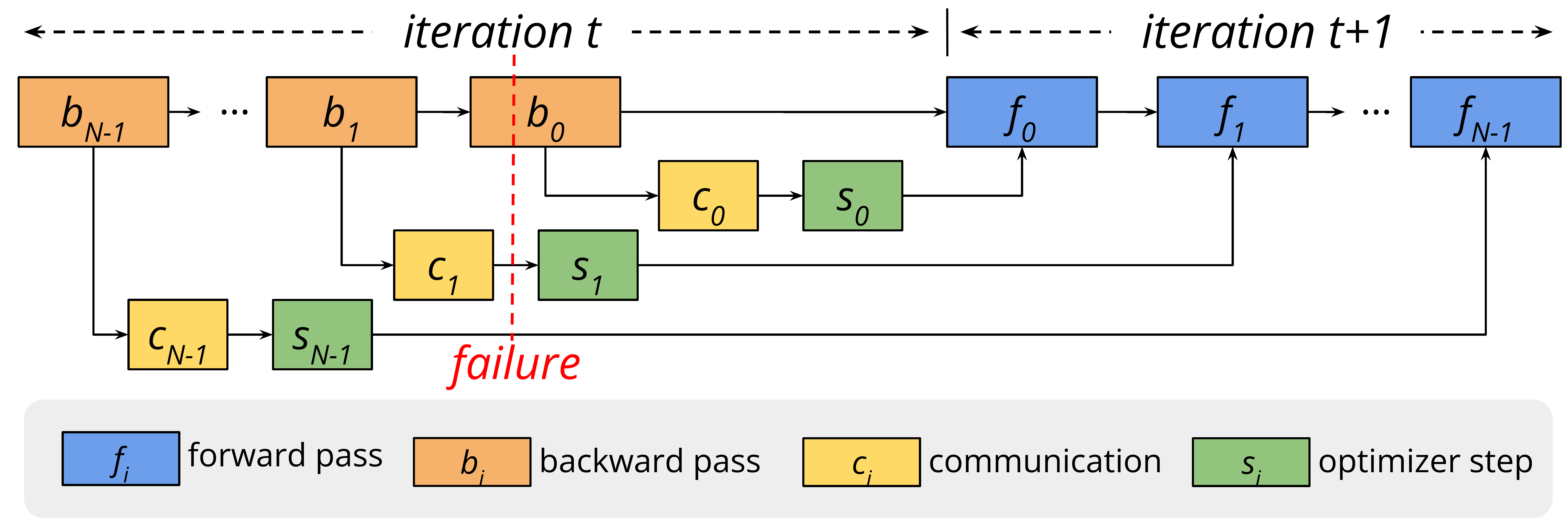}
    \caption{Crash-consistency problem in layer-wise wait-free update. The number is the layer index. Arrows represent dependency. The red dashed line indicates the failure.}
    \label{fig:layer_wise_update_overlap_with_backpropgation}
\end{figure}

\subsection{Crash-consistency Problem}
\label{sec2.3:crash_consistency problem}

Most DL frameworks~\cite{abadi2016tensorflow, paszke2019pytorch, chen2015mxnet, yuan2021oneflow} adopt wait-free model updates, as illustrated in Figure~\ref{fig:layer_wise_update_overlap_with_backpropgation}. Model state update of a DNN layer can be performed as soon as the gradient of that layer is ready. If a worker crashes during the update, the other workers are in an inconsistent state, where some layers are updated, and the others are not. The problem exists not only in data parallelism, but also in pipeline parallelism, where the DNN model is spread across multiple workers. Model state updates occur at different times due to the dependencies of the computation graph, and a worker moves on to the next training iteration once the part of the model state it hosts has been updated. Such inconsistencies can lead to a degradation of the accuracy of the final trained model~\cite{qiao2019fault, maeng2021understanding}.

Elastic Horovod solves the problem with the snapshot operation. However, it can be costly for large models (\S\ref{sec2.2:fault_tolerance_in_deep_learning}). Another workaround is to wait for gradients of all layers to be ready before updating the model state at the workers (by adding a barrier before the model update). With this method, the other workers can still complete their model updates and remain consistent even if a worker fails during the update. However, this update method incurs more waiting. In \S\ref{sec:undo_update}, we propose a better solution to tackle the crash-consistency problem without snapshotting or incurring the waiting.

\subsection{Logging-based Failure Recovery}
\label{sec2.4:logging_based_failure_recovery}

The logging method logs a job's application data at runtime and exactly replays the computation during failure recovery. Two main types of data are logged. Spark~\cite{zaharia2012resilient} and Ray~\cite{moritz2018ray, wang2019lineage} record lineage, i.e., the computation graph. Other systems record (or just buffer) raw, intermediate data~\cite{hwang2005high, shen2014fast}. For example, in \textit{upstream backup} of stream processing systems~\cite{hwang2005high}, the upstream machines preserve the data in the output queue while the downstream machines are processing them. 

In DNN training, the computation graph is usually fixed, and the execution time of a single operator in the computation graph is usually in the order of milliseconds. Recording the lineage of fine-grained operators adds significant overhead but does not benefit much since operators do not fail very often~\cite{abadi2016tensorflow}. Therefore, we consider logging intermediate data. The parallelism paradigm determines the communication operators and thus the data to be logged. Data parallelism and operator parallelism use collective communications, such as all-reduce and all-gather~\cite{zheng2022alpa}. Collective communications have complex data dependencies (e.g., many-to-many), thus complicating logging. Pipeline parallelism performs point-to-point communication, which simplifies logging. Also, the communication data volume in pipeline parallelism is much smaller than in data parallelism and operator parallelism for transformer models, and the current mainstream large models are usually transformer-based models~\cite{brown2020language, narayanan2021efficient}. Therefore, we advocate logging for pipeline parallelism. 

%% file: overview.tex
\section{\sysname{} Overview}
\label{sec:overview}

Given a distributed DNN training job and its specific parallelism configuration, \sysname{} introduces failure resilient mechanisms into the training job, targeting reduced failure recovery time without affecting the training throughput and final model accuracy. The distributed DNN training job spans a cluster of physical machines, with each machine hosting one or multiple workers. We focus on a fail-stop model~\cite{schlichting1983fail} throughout the paper, which is more common in real-world clusters~\cite{jeon2019analysis}. A machine in the cluster may crash during training, losing the volatile model states of workers it hosts, i.e., parameters and optimizer states which are mainly stored on the GPUs.

\sysname{} decides on a fault tolerance strategy before the training job starts. It always exploits redundancies if available (i.e., if the model state has at least one replica on another machine) because replication-based recovery achieves both low runtime and recovery overhead. When a replica is unavailable, and pipeline parallelism is used, and logging is \textit{worth doing} (\S\ref{sec:logging_use_case}), then use logging-based recovery. If none of the above conditions are met, use global checkpointing only. In any case, global checkpointing is performed periodically to ensure that the system remains on track in case of a catastrophic failure (e.g., loss of all replicas or logging data). Replication-based recovery and logging-based recovery can be combined to use, as hybrid parallelism is common for large DNN training, e.g., parts of the model use data parallelism while other parts use pipeline parallelism~\cite{zheng2022alpa}.

A machine failure can be detected by catching communication errors by workers that communicate with the failed machine. After the failure is detected, a replacement machine will be added to the training job. The surviving workers stop training and start the failure recovery procedure. Surviving workers first resolve the model state inconsistency issue (\S\ref{sec2.3:crash_consistency problem}) with the \textit{update-undo} approach (\S\ref{sec:undo_update}). Recovery is then performed for the replacement of the failed workers. For replication-based recovery, one of the surviving workers which holds the replica broadcasts the model state to the replacement workers (which uses data parallelism with the surviving worker). For logging-based recovery, replacement workers load the most recent checkpoint and then re-compute the lost iterations based on the logged data until recovering up to the pre-failure iteration. We also discuss multiple failures and cascading failures in Appendix~\ref{sec:multiple_cascading_failures}.

For the example in Figure~\ref{fig:3d_parallelism}, logging-based recovery can be used since replicas are unavailable and pipeline parallelism is used across the two machines. We record data of inter-machine communication during training (\S\ref{sec:4.1:basic_mechanism}): GPU 3 \& 7 log the intermediate activations in the forward pass, while GPU 11 \& 15 log the gradients in the backward pass.

%% file: system_design.tex
\section{Update-undo}
\label{sec:undo_update}

\begin{table}[t]
 \centering
 \caption{Operators used in five representative optimizers. EW = element-wise; Inv. = invertible.}
 \begin{tabular}{|c|c|c|c|c|c|c|}
  \hline
  \multicolumn{2}{|c|}{\scriptsize{Operator}} & \scriptsize{SGD} & \scriptsize{\makecell[c]{Adam\\~\cite{kingma2014adam}}} & \scriptsize{\makecell[c]{AdamW\\~\cite{loshchilov2017decoupled}}} & \scriptsize{\makecell[c]{LAMB\\~\cite{you2019large}}} & \scriptsize{\makecell[c]{AMSGrad\\~\cite{reddi2019convergence}}} \\
  \cline{1-7}
  \multirow{5}*{\footnotesize{Inv.}} & \footnotesize{EW add} & \checkmark & \checkmark & \checkmark & \checkmark & \checkmark \\
  \cline{2-7}
  ~ & \footnotesize{scalar mul} & \checkmark & \checkmark & \checkmark & \checkmark & \checkmark \\
     \cline{2-7}
  ~ & \footnotesize{EW mul} & ~ & \checkmark & \checkmark & \checkmark & \checkmark \\
  \cline{2-7}
  ~ & \footnotesize{EW sqrt} & ~ & \checkmark & \checkmark & \checkmark & \checkmark \\
  \cline{2-7}
  ~ & \footnotesize{EW div} & ~ & \checkmark & \checkmark & \checkmark & \checkmark \\
  \hline
  \multirow{2}*{\footnotesize{\makecell[c]{Not\\~}}} & \footnotesize{EW-max} & ~ & ~ & ~ & ~ & \checkmark \\
  \cline{2-7}
  \footnotesize{Inv.} & \footnotesize{sum} & ~ & ~ & ~ & \checkmark & ~ \\
  \hline
 \end{tabular}
 \label{tab:invertible_operations}
\end{table}

We propose undoing the update to address the crash-consistency problem (\S\ref{sec2.3:crash_consistency problem}). Our idea is simple: if a failure occurs during the model update when some parameters at the workers have been updated and some have not, the surviving workers will \textit{undo} the update for the updated parameters. In addition to model parameters, optimizer states, such as the momentum, also need to be restored. Finally, all workers return to a consistent version of the model state.

\begin{algorithm}[t]
    \caption{SGD with Momentum}
    \label{alg:sgdm}
    \begin{algorithmic}[1]
\State \textbf{Input:} learning rate sequence $\{\eta_t\}^{T}_{t=1}$; weight decay $\lambda > 0$; momentum parameter $0\leq \mu \leq 1$; dampening for momentum $0\leq \tau \leq 1$.
\State \textbf{Initialize:} $x_1 \in \mathbb{R}^d$; $m_0 = 0 \in \mathbb{R}^d$.

\State \textbf{for} $i=1$ \textbf{to} $T$ \textbf{do}
    \State \hspace*{\algorithmicindent} $g_t = \nabla f(x_t)$
    \State \hspace*{\algorithmicindent} $m_t = \mu m_{t-1} + (1-\tau) (g_t + \lambda x_t)$
    \State \hspace*{\algorithmicindent} $x_{t+1} = x_t - \eta_t m_t$
\State \textbf{end for}
    \end{algorithmic}
\end{algorithm}

\begin{algorithm}[t]
    \setlength{\belowdisplayskip}{3pt}
    \caption{Undo SGD with Momentum}
    \label{alg:undo-sgdm}
    \begin{algorithmic}[1]
    \State \textbf{Input:} learning rate $\eta_t$; weight decay $\lambda > 0$; momentum parameter $0\leq \mu \leq 1$; dampening for momentum $0\leq \tau \leq 1$; $x_{t+1} \in \mathbb{R}^d$; $g_t \in \mathbb{R}^d$; $m_t \in \mathbb{R}^d$.  
    \State  $x_t = x_{t+1} + \eta_t m_t$
    \State $ m_{t-1} = (m_t -  (1-\tau) (g_t + \lambda x_t)) / \mu$
    \end{algorithmic}
\end{algorithm}

We observe that many update operators of optimizers are mathematically \textit{invertible}, i.e., for an operator $f$, there exists an inverse operator $f^{-1}$ that undoes the operation of $f$. For example, linear operators like element-wise addition and scalar multiplication are all invertible~\cite{harte2016invertibility}. Table~\ref{tab:invertible_operations} summarizes operators used in five representative optimizers. Algorithm~\ref{alg:undo-sgdm} demonstrates how the update of SGD with momentum (Algorithm~\ref{alg:sgdm}) can be undone (i.e., from $x_{t+1}$ to $x_{t}$ and from $m_{t}$ to $m_{t-1}$). More examples can be found in Appendix~\ref{sec:appendix_a}. If an optimizer only has linear operators, then undoing is straightforward. However, some optimizers involve non-linear operators, e.g., LAMB optimizer~\cite{you2019large} scales the gradients with the L2 norm of the parameters. For the LAMB optimizer, we can additionally save the L2 norm (a scalar), and recover the previous model state accordingly. For AMSGrad~\cite{reddi2019convergence}, update undo is not applicable. Although the undo algorithms are mathematically correct, the recovered state may slightly differ from the original state due to floating-point errors~\cite{gomez2017reversible}. Our experiments show that this minor error does not affect trained model accuracy (\S\ref{sec:e2e}).

\begin{figure}[t]
    \centering
    \includegraphics[width=\columnwidth]{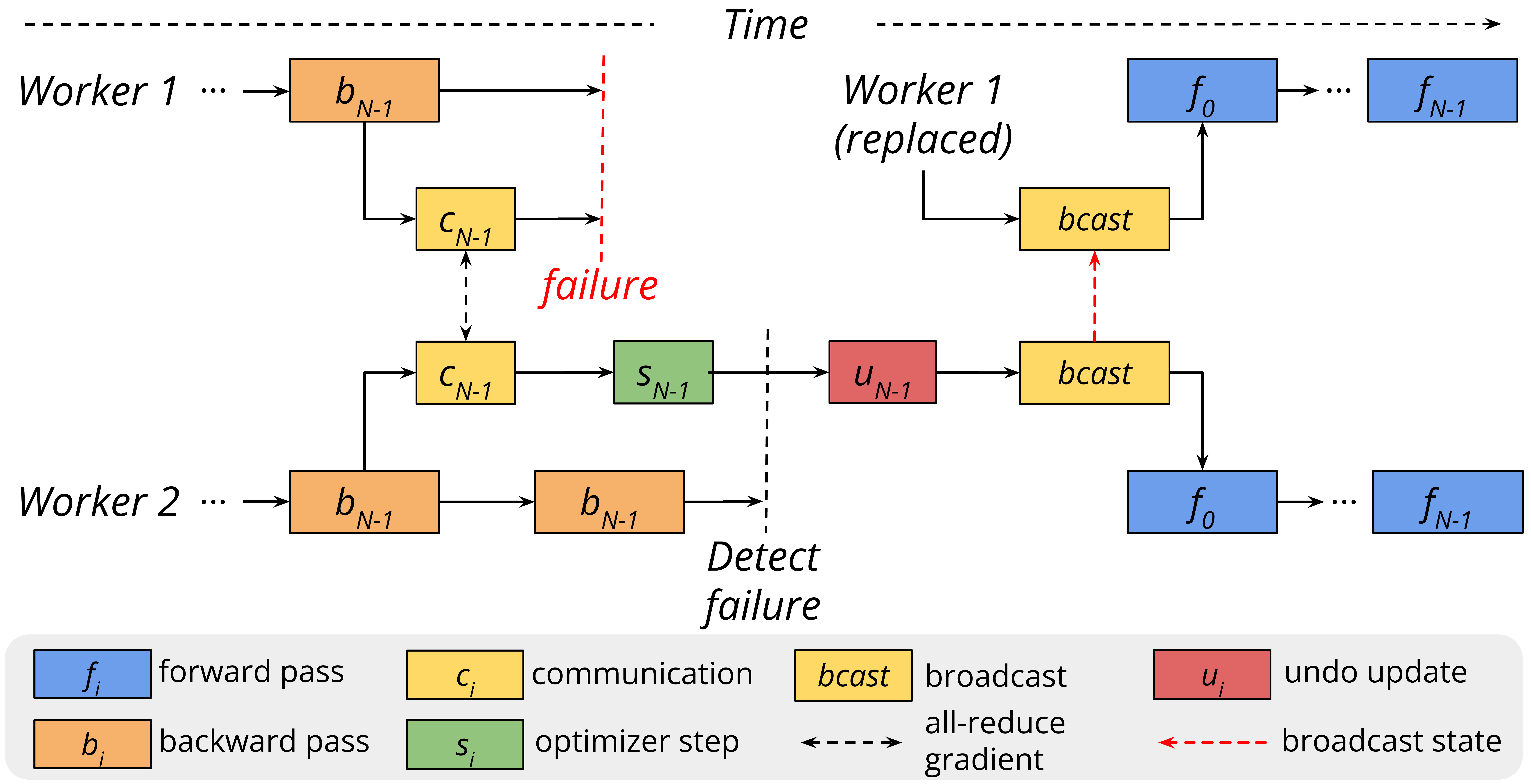}
    \caption{Replication-based recovery with data parallelism and undo operation.}
    \label{fig:undo}
\end{figure}

Figure~\ref{fig:undo} gives an example of how undoing updates helps replication-based recovery. Two workers train a DNN model using synchronous data parallelism. In training iteration $t$, worker 1 crashes and loses all the volatile states during the backward pass. At this time, worker 2 has already updated the parameters of layer $N-1$, but not the parameters of the other layers. Worker 2 then undoes layer $N-1$'s update to ensure consistency of its state. When the replacement of worker 1 joins the system, worker 2 sends its state to worker 1, and then both continue training from iteration $t$. 

Undoing the update does not require extra GPU memory. It only needs to cache the latest gradients $g_t$, a common practice in mainstream DL frameworks~\cite{paszke2019pytorch, abadi2016tensorflow, chen2015mxnet}. When gradients for the next iteration, $g_{t+1}$, are computed at the next backward pass, with synchronous training, parameters of all workers have already been updated by $g_t$, so that $g_{t+1}$ can safely overwrite $g_t$ in memory. It is only necessary to maintain one version of the gradients.

\section{Logging-based Recovery}
\label{sec:logging}

\subsection{Basic Mechanism}
\label{sec:4.1:basic_mechanism}

\vspace{1mm}
\noindent\textbf{What data to log.} For multi-GPU machines, GPUs are rare to fail individually, while a machine crash is more common~\cite{jeon2019analysis}. We hence do not record GPU-to-GPU communication within a machine but only \textit{inter-machine communication}, reducing the logging overhead substantially. The data to log include the intermediate activations in the forward pass and the gradients in the backward pass. In addition to saving the raw tensor, we need to record some metadata, including the sender and the receiver, and the timestamp (which contains identifiers of the current training iteration and the current micro-batch being trained). The timestamp is used to determine the order of the data to replay during recovery.

\vspace{1mm}
\noindent\textbf{How to log data.} 
Our logging method is similar to upstream backup (\S\ref{sec2.4:logging_based_failure_recovery}), i.e., the sender rather than the receiver logs the message. This way, the intermediate data needed for failure recovery are not lost but remain on the sender machines upon a failure. 
A sender does not need to store the message before sending it but can log it asynchronously in the background. A queue is set up for each worker. The worker keeps pushing outgoing tensors into the queue during training, while another background thread keeps reading tensors from the queue and doing the logging. Each worker flushes the queue of uncompleted logging tasks when detecting a failure in the training job. This \textit{asynchronous logging} significantly alleviates its impact on the training throughput.

Similar to asynchronous checkpointing (as in CheckFreq), which still incurs significant overhead (\S\ref{sec2.2:fault_tolerance_in_deep_learning}), we need to reduce the overhead of our asynchronous logging further. In synchronous pipeline-parallel training, there are many bubbles during which the worker is idle, which is an ideal time for logging (\S\ref{sec2.1:distributed_dnn_training}). With \textit{logging during the bubble time}, the outgoing tensor is not logged immediately after production, but waits until a bubble occurs. The waiting time is not longer than one training iteration because a bubble always exists within an iteration. Once the logging data is transferred to the CPU, the data on the GPU can be safely removed. Therefore, logging during the bubble time will not cause substantial data accumulation on the GPU. This way, logging is \textit{off the critical path}.

\begin{figure}[t]
  \centering
     \begin{subfigure}[b]{\columnwidth}
      \centering
    \includegraphics[width=\columnwidth]{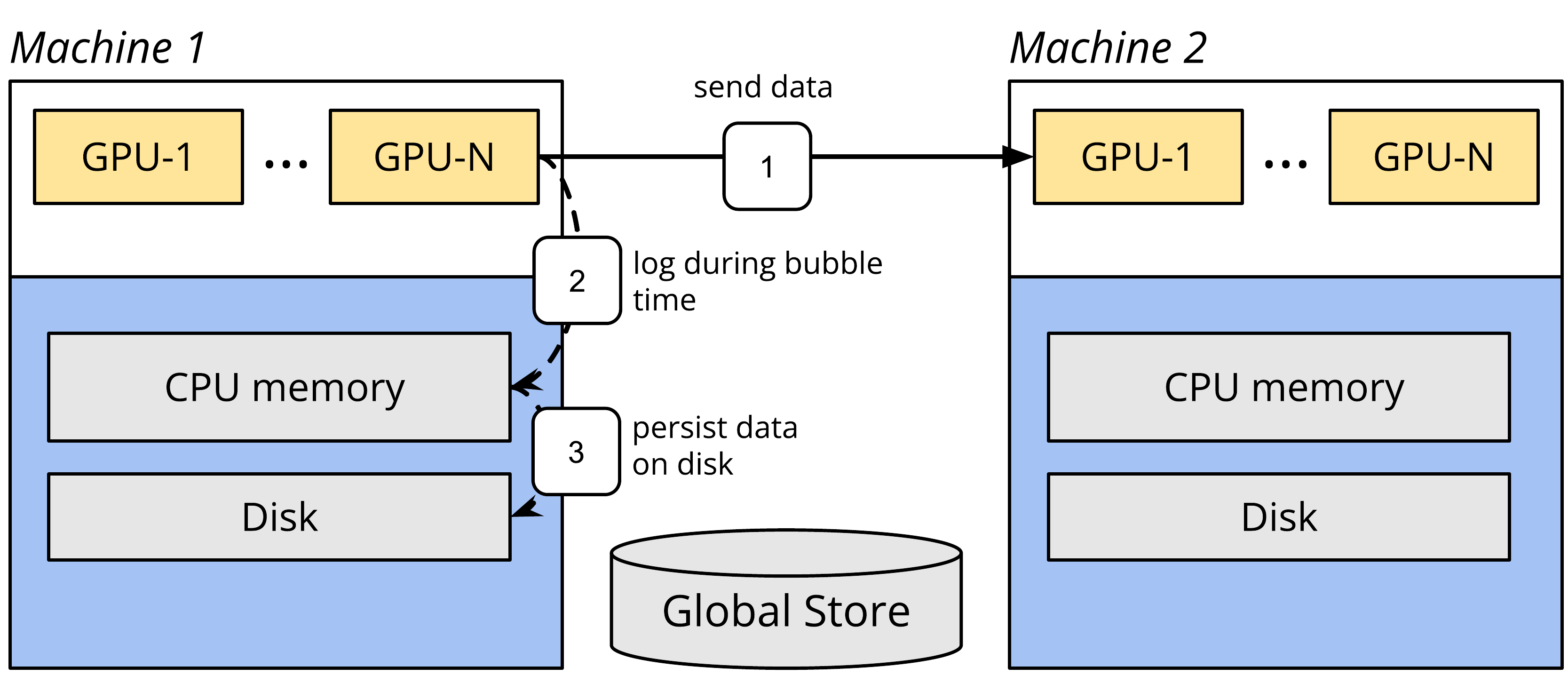}
    \caption{Logging during runtime.}
 \label{fig:logging_mechanism_runtime}
     \end{subfigure}
     \hfill
     \begin{subfigure}[b]{\columnwidth}
      \centering
    \includegraphics[width=\columnwidth]{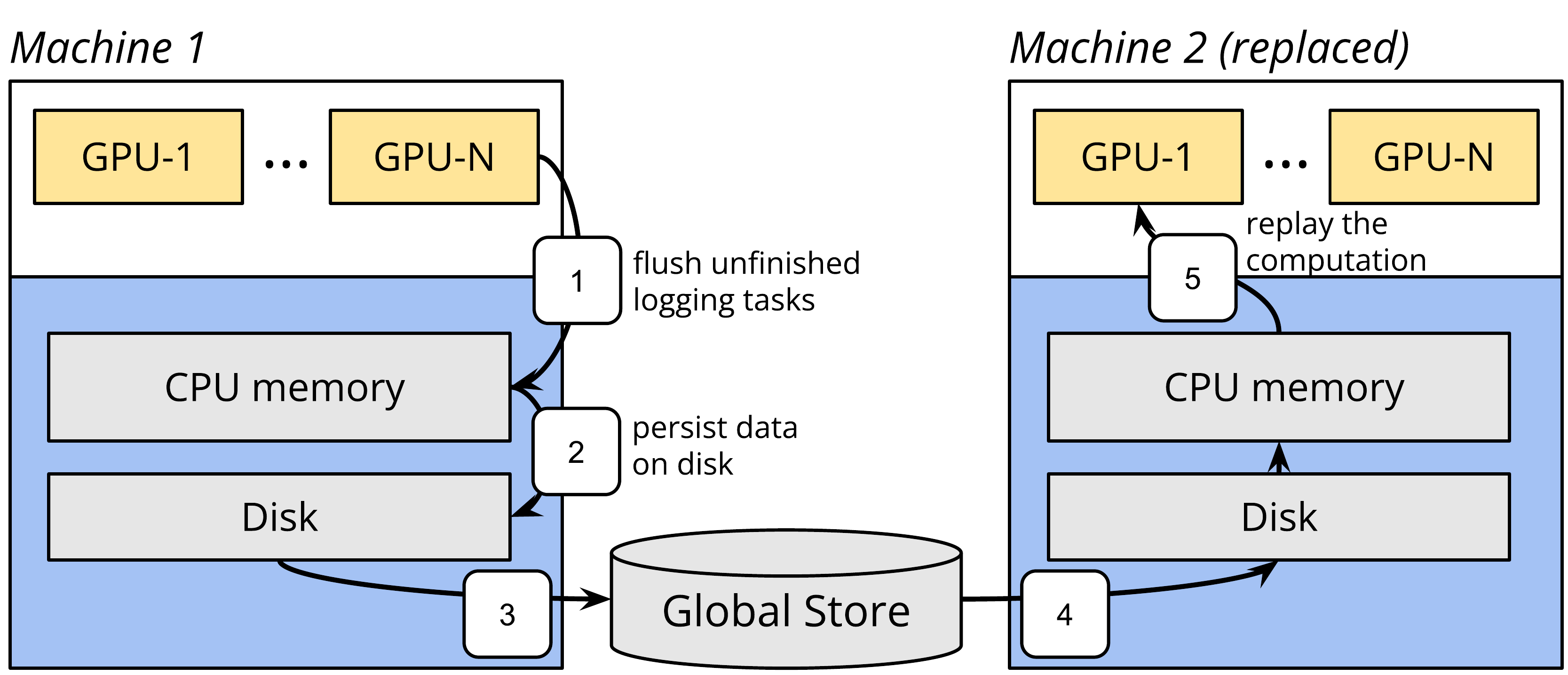}
    \caption{Logging during recovery.}
 \label{fig:logging_mechanism_recovery}
     \end{subfigure}
    \hfill
     \begin{subfigure}[b]{\columnwidth}
    \centering
    \includegraphics[width=\columnwidth]{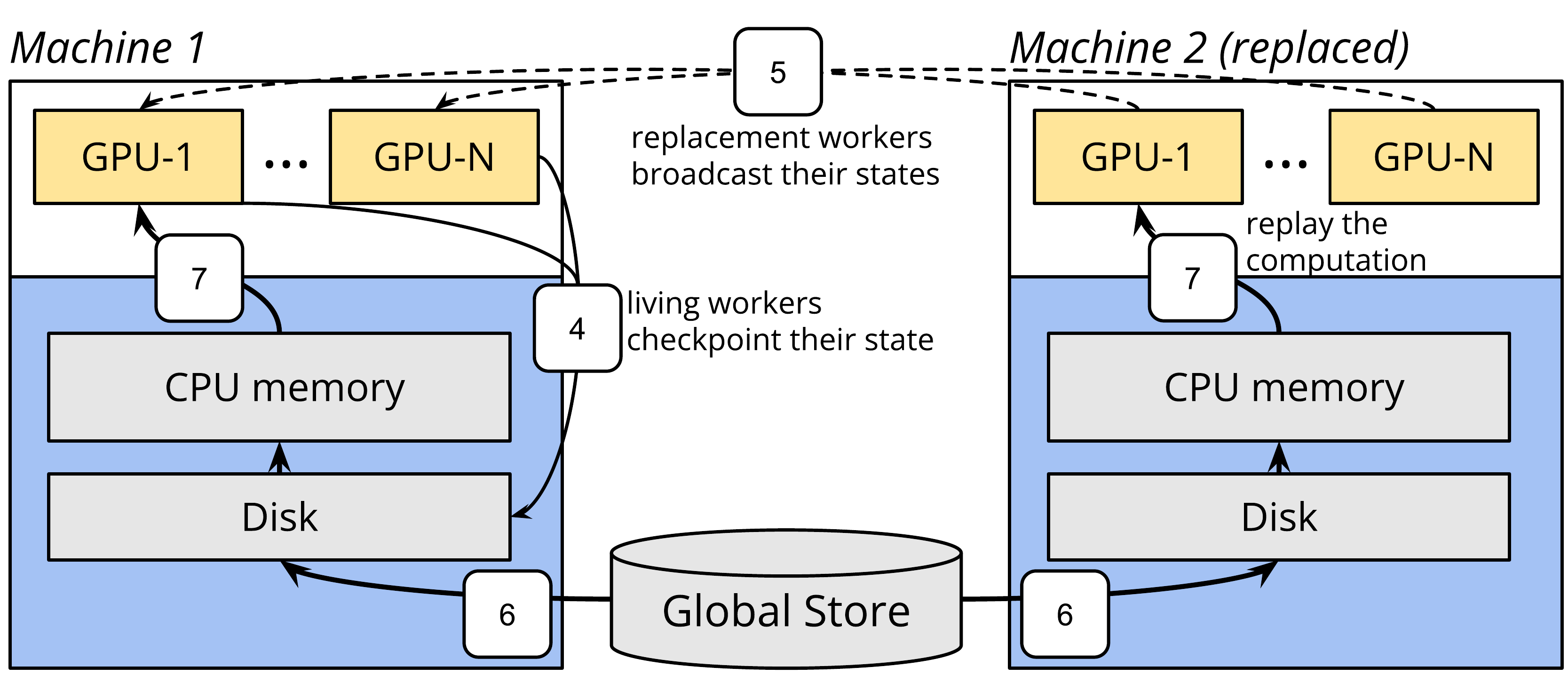}
    \caption{Logging with parallel recovery. Steps 1 to 3 are the same as in Figure~\ref{fig:logging_mechanism_recovery} and are omitted.}
     \label{fig:logging_mechanism_parallel_recovery}
    \end{subfigure}
    \caption{Logging mechanism.}
    \label{fig:logging_mechanism}
\end{figure}

Figure~\ref{fig:logging_mechanism_runtime} illustrates how logging is done during failure-free execution. After a worker (GPU) sends a tensor to a downstream worker (step 1), the data remains on the GPU for a short time. During the next bubble time, the tensor is copied asynchronously to the CPU memory (step 2). A background thread writes the logging data in CPU memory to the disk and saves them into a file (step 3).

\vspace{1mm}
\noindent\textbf{Recovery.} Figure~\ref{fig:logging_mechanism_recovery} shows how failure recovery can be made using the logging data. Suppose machine 2 crashes. The surviving upstream workers (machine 1 in this example) copy unlogged data to CPU memory (step 1) and then write them to the disk (step 2). The upstream workers then upload their logging files to global storage (step 3), e.g., HDFS~\cite{shvachko2010hadoop}. Machine 2's replacement downloads the logging files that it needs from the global store to its local disk (step 4), loads the latest checkpoint, and replays previously received tensors from the logging file in the exact order of their timestamps (step 5). The step 3, 4, and 5 can be executed in a pipeline by chunking the logging file into multiple smaller files. If necessary, the surviving workers will undo the update (\S\ref{sec:undo_update}) (not shown in the figure).

The most significant difference with pure global checkpointing is that surviving workers do not need to load the checkpoint and roll back their training progress for recovery. Only relaunched workers on the replacement machine do. The recovery scope is limited to the local computation graph on the failed machine rather than the whole computation graph, thus expediting the recovery.

\vspace{1mm}
\noindent\textbf{Garbage collection.} All earlier logging files are obsoleted after a global checkpointing, and garbage is collected because the system can directly load the latest checkpoint then. Even though the logging size increases as the number of iterations increases, the size is upper bounded due to periodic global checkpointing. Therefore, the frequency of global checkpointing determines the upper bound of the logging size. We will discuss more the storage overhead in \S\ref{sec:4.3:selective_logging}.

\vspace{1mm}
\noindent\textbf{Consistency.} 
We replay the computations in the same order as the pre-failure execution using timestamps, using the same inputs as the pre-failure computation. Note that logging requires the computation to be \textit{deterministic} (i.e., the same input leads to the same output). Otherwise, we would get different outputs when re-computing with the logged data. We provide the details of achieving determinism in \S~\ref{sec:impl}.

\subsection{Parallel Recovery}
\label{sec:4.2:parallel_recovery}
We utilize the surviving workers to assist in recovery of the failed workers. 
Since the intermediate results of all micro-batches have been logged, we can perform data-parallel training based on the logged data to expedite the re-computation of the lost states. Specifically, all workers, including replacement workers and surviving workers, retrieve the logging files and have a copy of the computation graph on the failed machine. Each worker reads logging data of different micro-batches from the logging files and uses them as input to re-compute gradients, synchronizes gradients with other workers computing other micro-batches and then performs the model update. This way, the micro-batches are re-computed in parallel by multiple workers for the failed machine, accelerating recovery while ensuring logical equivalence to executing these micro-batches sequentially by each replacement worker. If a batch is divided into $m$ micro-batches and we use $d$ workers for parallel recovery, each worker is assigned with $m/d$ micro-batches for re-computation.

\begin{figure}[t]
    \centering
    \includegraphics[width=\columnwidth]{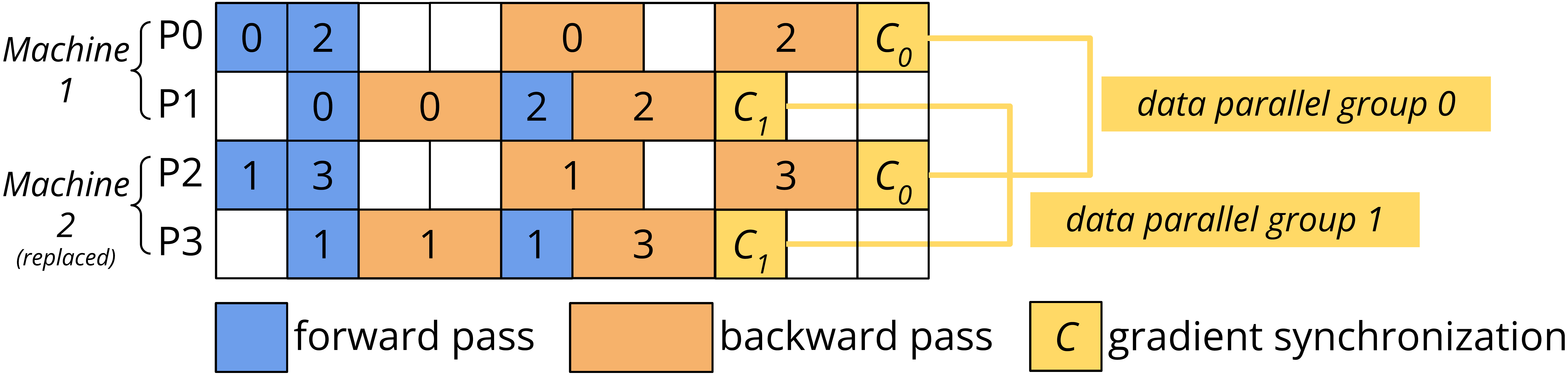}
    \caption{Parallel recovery. The number of micro-batches is 4 ($m=4$) and we use 2 workers ($d=2$) in each data-parallel recovery group. Each machine has two workers. Suppose machine 2 crashes and is replaced. We decompose the pipeline in Figure~\ref{fig:pipeline_parallelism_1f1b} into data-parallel 2-stage sub-pipelines.}
    \label{fig:pipeline_parallelism_parallel_recovery}
\end{figure}

An example is given in Figure~\ref{fig:pipeline_parallelism_parallel_recovery}, where two machines run a 4-stage training pipeline. Machine 2 (hosting stages 2 and 3) fails, and Machine 1 (hosting stages 0 and 1 in normal training) assists in the recovery computation of the replacement machine.
Worker P0 on Machine 1 and worker P2 on the replacement machine re-compute the stage-2 model in a data-parallel manner, each using two micro-batches (0, 2 and 1, 3, respectively); worker P1 
and worker P3 
re-compute the stage-3 model, using micro-batches 0, 2 and 1, 3, respectively.
Note that extra time is needed for gradient synchronization with parallel recovery. 

The parallel recovery procedures are given in Figure~\ref{fig:logging_mechanism_parallel_recovery}. 
Similar to Figure~\ref{fig:logging_mechanism_recovery}, at a surviving worker, uncommitted logging data are first flushed and uploaded to the global store (steps 1 to 3, omitted in the figure). Then the surviving workers checkpoint their states (step 4). The replacement workers load their model states from their latest checkpoints and broadcast their states to the surviving workers (step 5). Meanwhile, all workers download logging files from the global store (step 6) and select the logging data of corresponding micro-batches for re-computation 
(step 7). After the recovery, the surviving workers load their checkpoints to restore their original model parameters and optimizer states (not shown in the figure). 

\subsection{Selective Logging}
\label{sec:4.3:selective_logging}

Logging all cross-machine messages may consume large storage space. We next investigate a trade-off between the storage space and the recovery time with selective logging. 
Our idea is to group machines and log inter-group communication but not intra-group communication. We can consider the original approach as a particular case, where each machine forms a group. 
In this way, if one machine in a group fails, training on the entire group of machines needs to be rolled back from the latest checkpoint, as we do not record intra-group communication. As a result, 
and the recovery time will be longer. 
Thus, selective logging trades recovery time for space overhead.
A simple grouping strategy is to have a balanced number of machines in each group. However, due to the often unbalanced model partition in pipeline parallelism~\cite{fan2021dapple}, this grouping strategy is usually suboptimal. 
Given a storage capacity constraint for logging data, how do we group machines to minimize the failure recovery time? 
Suppose we have $N$ machines and create $N$ groups initially. We profile the averaged per-iteration computation time $R(G_i)$ for each group $G_i$. For each pair of adjacent groups $G_i$ and $G_{i+1}$ (i.e., hosting adjacent workers in the pipeline), we obtain the transmission size per iteration $M(G_i, G_{i+1})$ between them. 
Then with storage capacity limit $M_{\text{max}}$, network bandwidth $B$ (assuming homogeneous bandwidth) and checkpointing interval $T$ (iterations), we aim at finding a group configuration $\mathcal{G} = \{G_1,\ldots, G_k\}$ that minimizes the overall recovery time $R$: 
{\small
\begin{equation}
    \min_{\mathcal{G}} R(\mathcal{G})\quad \text{s.t.}\; M(\mathcal{G}) \leq M_{\text{max}}, \nonumber
\end{equation}
}
\noindent where $M(\mathcal{G})$ denotes the 
overall storage space needed by the logging data. As discussed in \S\ref{sec:4.1:basic_mechanism}, it is determined by the global checkpointing frequency:  
{\small
\begin{equation}
    M(\mathcal{G}) = T\cdot \sum_{G_{i}, G_{i+1}\in \mathcal{G}} M(G_i, G_{i+1}). \nonumber 
\end{equation}
}
\noindent Suppose we merge two adjacent groups $G_i$ and $G_{i+1}$, and have the following recovery time for the merged group: 
{\small
\begin{equation}
    R(G_i, G_{i+1}) = R(G_i) + R(G_{i+1}) + M(G_i, G_{i+1}) / B, \nonumber
\end{equation}
}
\noindent where $M(G_i, G_{i+1}) / B$ is the point-to-point communication time between the two adjacent groups. 
We ignore the bubble time for simplicity, and derive the change in overall recovery time $R$ and overall space overhead $M$: 
{\small
\begin{align} 
    \Delta R &= {} R(G_i, G_{i+1})\cdot \frac{|G_i| + |G_{i+1}|}{N} - R(G_i)\cdot \frac{|G_i|}{N} \nonumber  \\
    &\quad - R(G_{i+1})\cdot \frac{ |G_{i+1}|}{N}, \nonumber
\\
\Delta M &= {}
M(G_i, G_{i+1})\cdot  T, \nonumber
\end{align}
}

\noindent where $|G_i|$ is the number of machines in $G_i$. $\Delta R$ is calculated assuming that each machine has an equal failure probability. 
Note that $\Delta R$ is always positive. We minimize increased recovery time $\Delta R$ per unit storage space reduction 
when merging $G_i$ and $G_{i+1}$, i.e., 
minimize $\Delta R / \Delta M$. 

To identify the grouping of machines, we iteratively merge two adjacent groups with the smallest $\Delta R / \Delta M$, until the overall space consumption is less than $M_{\text{max}}$. 
Note that it runs for at most $N-1$ iterations, at which point all machines form a single group and there will be no logging. So the time complexity is at most $O(N^2)$. 

If parallel recovery is used, the recovery of a group $G_i$ is parallelized by at most $\lfloor N /|G_i| \rfloor$ data-parallel groups. For simplicity, we assume it can achieve linear scalability with data parallelism. Thus, 
we divide the $R(G_i)$ with $\lfloor N /|G_i| \rfloor$ in calculation.

\subsection{Use Case}
\label{sec:logging_use_case}
Not all cases are suitable for logging. For example, it would be better to checkpoint a model when the logging size far exceeds the model size. Typically, the intermediate activations for CNN-based models would be massive and unsuitable for logging (even unsuitable for pipeline parallelism)~\cite{rhu2016vdnn}. We can calculate the per-iteration logging size. For transformer-based models, the intermediate activation/gradient size would be $\text{micro\_batch\_size} \times \text{hidden\_size} \times \text{sequence\_length}$ in a micro-batch~\cite{zero-infinity}. Further, we can calculate the bubble time ratio according to the pipeline schedule (\S\ref{sec2.1:distributed_dnn_training}). Given the iteration time and PCIe bandwidth, we can determine whether the logging data can be transferred from GPU to CPU within the bubble time. If not, then logging is not worth doing.

%% file: implementation.tex
\section{Implementation}
\label{sec:impl}

We implement \sysname{} in PyTorch 1.9.0~\cite{paszke2019pytorch} with NCCL 2.7.6~\cite{jeaugey2017nccl}, using 2.6k LoC in Python. We also add about 400 lines of C++ code for PyTorch and NCCL.

\vspace{1mm}
\noindent\textbf{Failure detection.} 
We launch a background thread on each worker that uses NCCL's \texttt{ncclCommGetAsyncError()} function to keep polling whether a communication failure has occurred. If a failure occurs, the worker first sets a failure flag to true in a global key-value store and then aborts its own NCCL communicators. The global key-value store is co-located with the master machine (rank 0). Other workers' background threads also poll this flag from the global key-value store, and if a worker finds that the flag is set to true, it will abort its own NCCL communicators.

\vspace{1mm}
\noindent\textbf{Update-undo.} In data parallelism, we insert a CUDA event after the all-reduce operation for each tensor and query whether it has been completed before updating the gradient's corresponding parameter. If it does not complete, it waits until the all-reduce operation completes. After it completes, the CUDA kernels for the corresponding parameter are launched to update the parameter and optimizer states, and the parameter is marked as updated. Note that even if there is a failure at this point, we need to let these kernels finish executing. Upon a failure, surviving workers undo the update of parameters that are marked updated.

In pipeline parallelism, the model parameters on different stages are updated at different points in time due to computational dependency. Therefore, surviving  workers need to exchange their current iteration number to determine the consensus pre-failure iteration after a failure occurs. Workers with a greater iteration number than the consensus pre-failure iteration need to undo the update.

\vspace{1mm}
\noindent\textbf{Logging.} We use a dedicated CUDA stream to copy logging data from the GPU to the CPU for asynchronous logging. We insert a CUDA event after the copy operation to check if the copy operation is completed. After the main thread launches asynchronous copying operations at bubble time, it sends the CUDA event with the corresponding tensor and metadata to a queue. A background thread keeps reading items from the queue and checks if the asynchronous copy is complete by checking the CUDA event status. If completed, the thread saves the data to a file. For the global store described in \S\ref{sec:4.1:basic_mechanism}, we support HDFS~\cite{shvachko2010hadoop} and Amazon S3.

\vspace{1mm}
\noindent\textbf{Determinism in Logging.}
Nondeterminism in DNN training may come from the random number seeds and algorithms themselves. For example, in DNN training, some convolution algorithms in cuDNN are nondeterministic\footnotemark{}.\footnotetext{cuDNN reproducibility: \url{https://docs.nvidia.com/deeplearning/cudnn/developer-guide/index.html\#reproducibility}} 
We set \texttt{torch.backend.cudnn.deterministic=True} to resolve this issue. In addition, for convolutional operations, PyTorch also benchmarks multiple algorithms in the first run, selects the fastest one, and caches this choice so that the same algorithm can be directly selected later. However, there are still some slight differences in the computational results of different deterministic algorithms for the same input. In order to ensure that the worker selects the same convolutional algorithm after failure recovery as before failure, we save the previous benchmark results for failure recovery.

\vspace{1mm}
\noindent\textbf{Usage.} We provide an easy-to-use interface for users. A user only needs to provide a user-defined function (UDF) to train for one iteration and specify fault tolerance and training configurations. Then fault tolerance is in place during the user's model training, and recovery upon a failure can be automatically run without requiring user involvement. 

%% file: evaluation.tex
\section{Evaluation}

\begin{figure*}[t]
    \centering
    \begin{subfigure}[b]{0.65\columnwidth}
        \centering
        \includegraphics[width=\columnwidth]{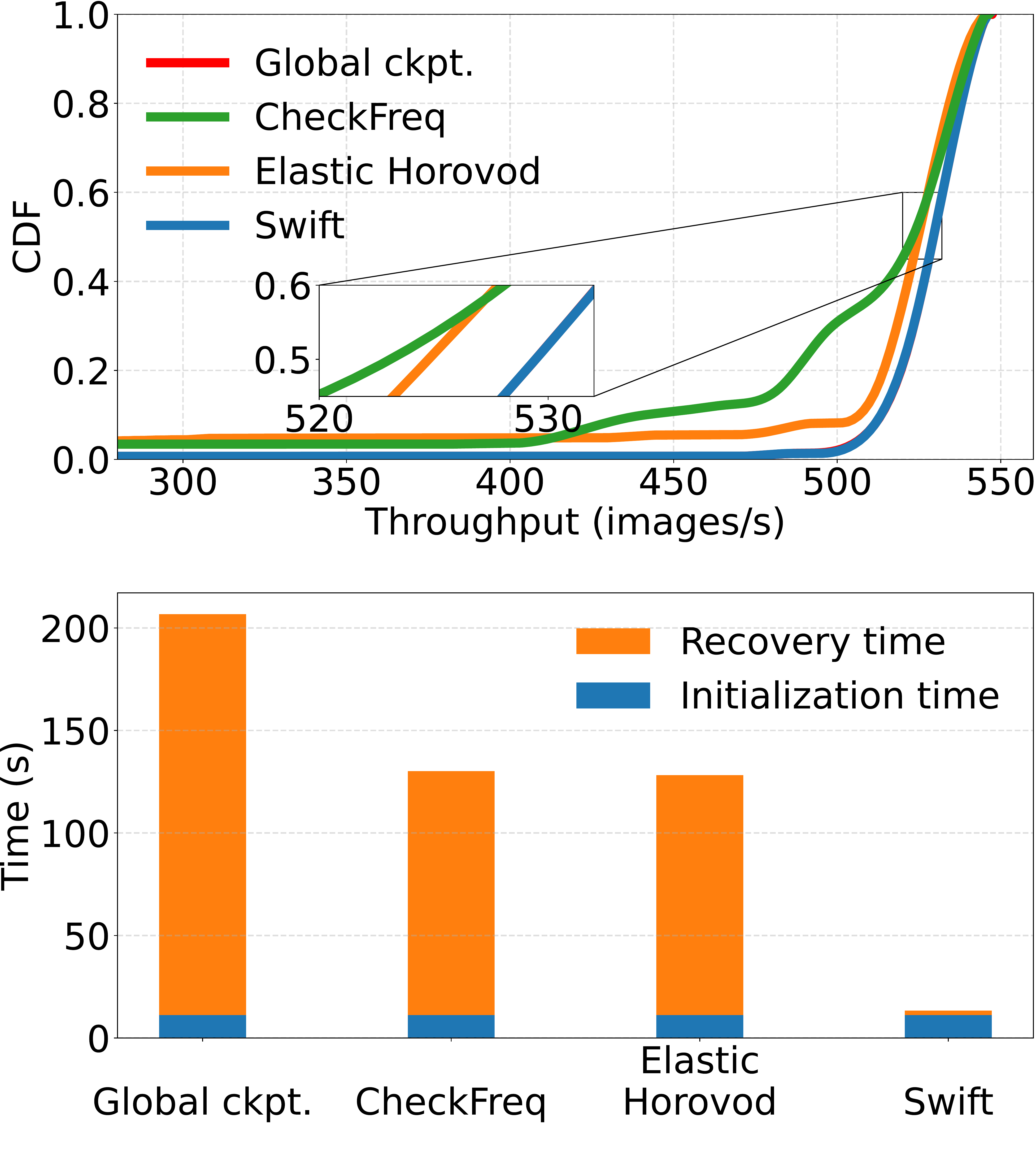}
        \caption{Wide-ResNet-50}
        \label{fig:micro_benchmark_speedup_wide_resnet}
    \end{subfigure}
    \hfill
    \begin{subfigure}[b]{0.65\columnwidth}
        \centering
        \includegraphics[width=\columnwidth]{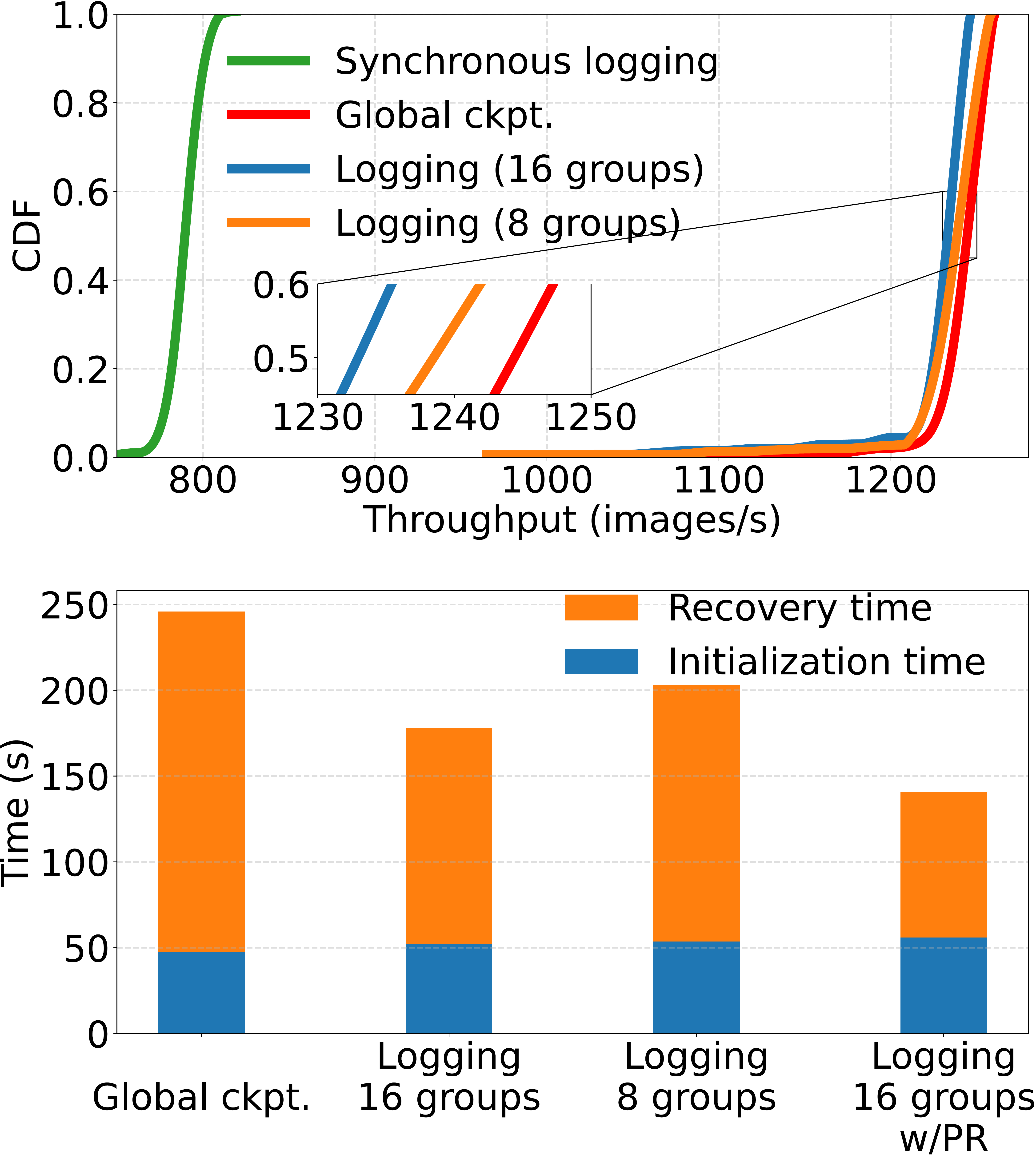}
        \caption{ViT-128/32}
        \label{fig:micro_benchmark_speedup_vit}
    \end{subfigure}
    \hfill
    \begin{subfigure}[b]{0.65\columnwidth}
        \centering
        \includegraphics[width=\columnwidth]{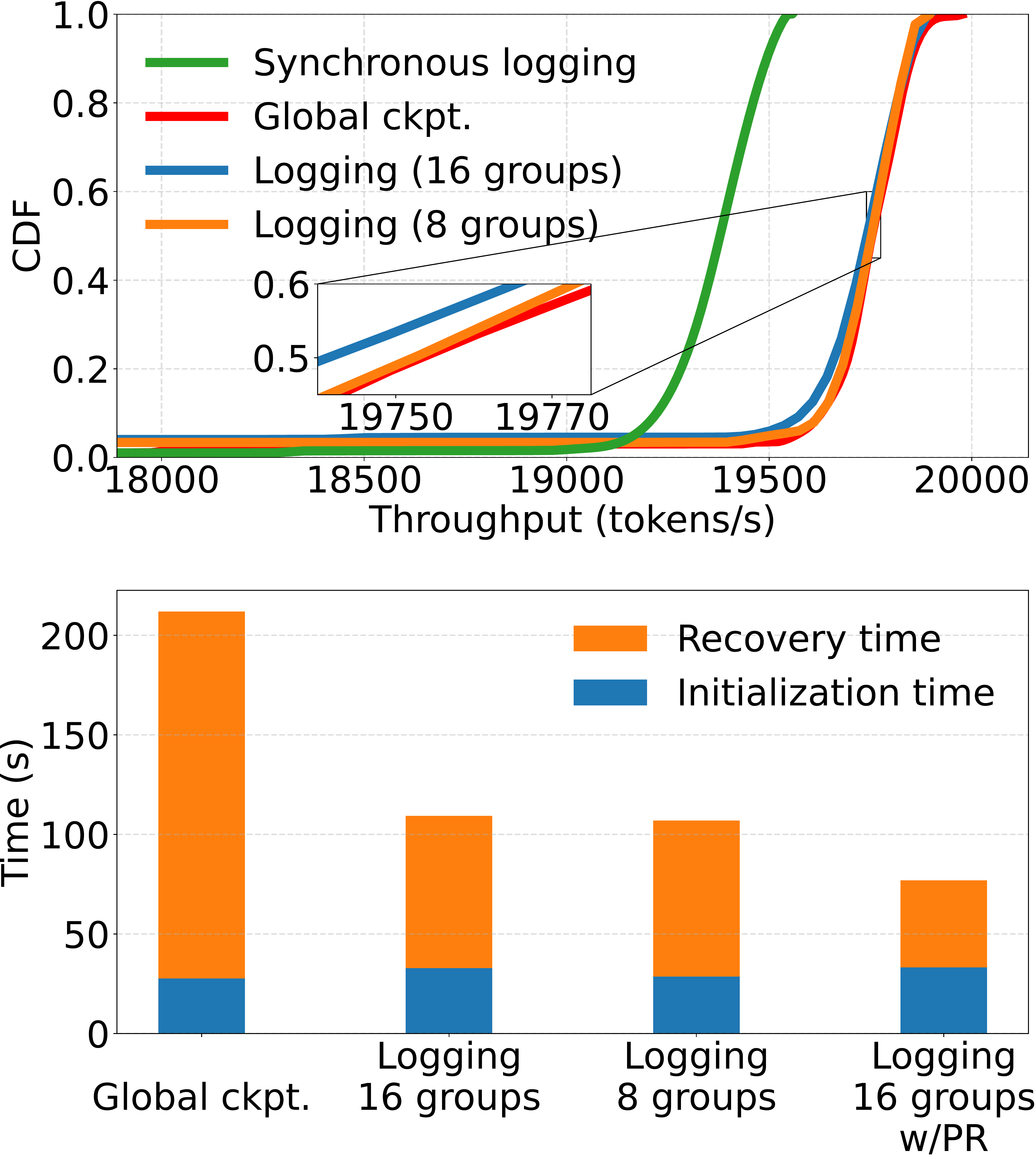}
        \caption{BERT-128}
        \label{fig:micro_benchmark_speedup_bert}

    \end{subfigure}
    \caption{Failure-free training throughput (top) and recovery time (bottom). PR = parallel recovery.}
    \label{fig:micro_benchmark_speedup}
\end{figure*}

\label{sec:experimental_setup}
\noindent\textbf{Testbed.} We experiment on 16 DGX-2 machines, each equipped with eight 32 GB Tesla V100 GPUs (NVLink interconnect), 160 CPUs, 1.5 TB memory, and 3.6 TB NVMe SSD disks. The machines are connected via 40Gbps Ethernet. We build an HDFS cluster on these machines as the global storage.

\begin{table}[t]
    \centering
    \caption{Benchmark Models. DP = data parallelism. PP = pipeline parallelism.} 
    \begin{tabular}{ccccc}
        \toprule
        \multirow{2}{*}{\footnotesize{Model}} &  \multirow{2}{*}{\footnotesize{Dataset}} & \footnotesize{Batch} & \footnotesize{\#params} & \multirow{2}{*}{\footnotesize{Parallelism}} \\
         &  & \footnotesize{size} & \footnotesize{(billion)} & \footnotesize{} \\
        \midrule
        \footnotesize{Wide-ResNet-50} &\footnotesize{ImageNet~\cite{russakovsky2015imagenet}}
        & \footnotesize{256} & \footnotesize{1.23} &\footnotesize{DP} \\
        \footnotesize{ViT-128/32} &\footnotesize{ImageNet}
        & \footnotesize{4096} & \footnotesize{1.64} &\footnotesize{PP} \\
        \footnotesize{BERT-128} &\footnotesize{Wikipedia~\cite{devlin2018bert}}
        & \footnotesize{512} & \footnotesize{1.11} & \footnotesize{PP} \\
        \bottomrule
    \end{tabular}
    \label{tab:benchmarks}
\end{table}

\vspace{1mm}
\noindent\textbf{Benchmark Models.} We evaluate \sysname{} on training large image classification and language models with billions of parameters, as given in Table~\ref{tab:benchmarks}. We scale up the original models in their respective papers: for Wide-ResNet-50~\cite{zagoruyko2016wide}, we increase the base channel size from 64 to 320; for BERT-Large~\cite{devlin2018bert} and ViT-Large/32~\cite{dosovitskiy2020image}, we increase the number of transformer layers from 24 to 128, keep the hidden size unchanged and refer to the enlarged models as BERT-128 and ViT-128/32, respectively. We use data parallelism to train Wide-ResNet-50 on two machines and four GPUs on each. To train ViT-128/32 or BERT-128 (with a maximum sequence length of 128), we use a 128-stage pipeline on all 16 machines, with each transformer layer occupying one GPU. We use SGD with momentum for Wide-ResNet-50 and ViT-128/32, and Adam for BERT-128~\cite{dosovitskiy2020image, dosovitskiy2020image, devlin2018bert}. We select the micro-batch number to maximize the performance, using 16 and 4 for ViT-128/32 and BERT-128, respectively. 
\sysname{} applies replication-based recovery to Wide-ResNet-50, and logging-based recovery to ViT-128/32 and BERT-128 (by default 16 machine groups and 8 machine groups in selective logging). We run each experiment for 200 iterations, perform a global checkpoint at the beginning of iteration 100, and kill a machine (rank 1) at the beginning of iteration 150.

\vspace{1mm}
\noindent\textbf{Baselines.} We compare \sysname{} with global checkpointing (default in PyTorch), CheckFreq~\cite{mohan2021checkfreq} and Elastic Horovod~\cite{elastichorovod}. We use CheckFreq's open-sourced code~\cite{checkfreq-code} and replace Elastic Horovod's snapshot implementation with CheckFreq's since it does not implement snapshotting to the CPU. We calculate the optimal snapshot frequency (once per 30 iterations) based on the algorithm suggested by CheckFreq and using the same permissible checkpoint overhead (3.5\%) as in CheckFreq's experiments. For logging-based recovery, we only compare with global checkpointing, as its checkpointing overhead is already very low (checkpointing is pipelined in pipeline-parallel training), and the performance of CheckFreq would be similar. Elastic Horovod is not applicable since it only supports data parallelism. We also introduce a synchronous logging method (calling \texttt{torch.save()} before sending a tensor) as a baseline to evaluate the effect of our asynchronous logging and logging during bubble time (\S\ref{sec:4.1:basic_mechanism}).

\vspace{1mm}
\noindent\textbf{Metrics.} We evaluate the training throughput and iteration time during failure-free execution, throughput during recovery, and recovery time. Training throughput is calculated by the number of images (or tokens) processed by all workers per training iteration. Initialization time counts from when workers detect the failure to when the replacements of failed workers join the training job. Recovery time is the duration from when the replacements of workers join the training job to the time they recover to the pre-failure iteration.

\subsection{Macro-benchmarks} \label{sec:micro_benchmarks}

\vspace{1mm}
\noindent\textbf{Replication-based recovery.}
Figure~\ref{fig:micro_benchmark_speedup_wide_resnet} shows that \sysname{}'s replication-based recovery incurs less runtime overhead than state-of-the-art methods during failure-free training. Training throughput of CheckFreq and Elastic Horovod degrades compared to the normal training (without any checkpoint or snapshot).

Figure~\ref{fig:micro_benchmark_speedup_wide_resnet} also presents the recovery time upon a machine failure at iteration 150. Global checkpointing takes a long time to recover, as all workers must load the checkpoint and re-compute the lost iterations (50 iterations in this experiment). CheckFreq and Elastic Horovod do frequent snapshotting but still need to re-compute 30 iterations (the last snapshot was captured at iteration 120). With \sysname{}'s replication-based recovery, surviving workers resolve inconsistencies by undoing updates and then broadcast replicas to the replacement workers. It reduces the recovery time by 98.9\%, 98.1\%, and 98.1\% as compared to global checkpointing, CheckFreq, and Elastic Horovod, respectively.

\vspace{1mm}
\noindent\textbf{Logging-based recovery.} Figure~\ref{fig:micro_benchmark_speedup_vit} and Figure~\ref{fig:micro_benchmark_speedup_bert} show that \sysname{}'s logging is slightly slower than global checkpointing during failure-free training of ViT-128/32 and achieves similar throughput for BERT-128. The slight delay for ViT-128/32 is because we use a large batch (4096) when training and the logging data size is relatively large. Synchronous logging significantly degrades training throughput, especially when training ViT-128/32, due to logging more data than BERT-128. \sysname{}'s asynchronous logging and logging during bubble time (\S\ref{sec:4.1:basic_mechanism}) take logging off the critical path, leading to a similar throughput compared to global checkpointing. 

Figure~\ref{fig:micro_benchmark_speedup_vit} and Figure~\ref{fig:micro_benchmark_speedup_bert} show that the recovery time with logging is substantially smaller than global checkpointing. With 16 machine groups, the recovery time is reduced by 36.0\% and 58.5\% for ViT-128/32 and BERT-128, respectively. This is because only the 8-stage sub-pipeline on the failed machine needs to be recovered, compared to re-running the whole 128-stage pipeline when using global checkpointing. Note that logging needs slightly more initialization time because it requires additional initialization operations such as creating a CUDA stream and logging threads

\begin{figure}[t]
    \centering
    \includegraphics[width=\columnwidth]{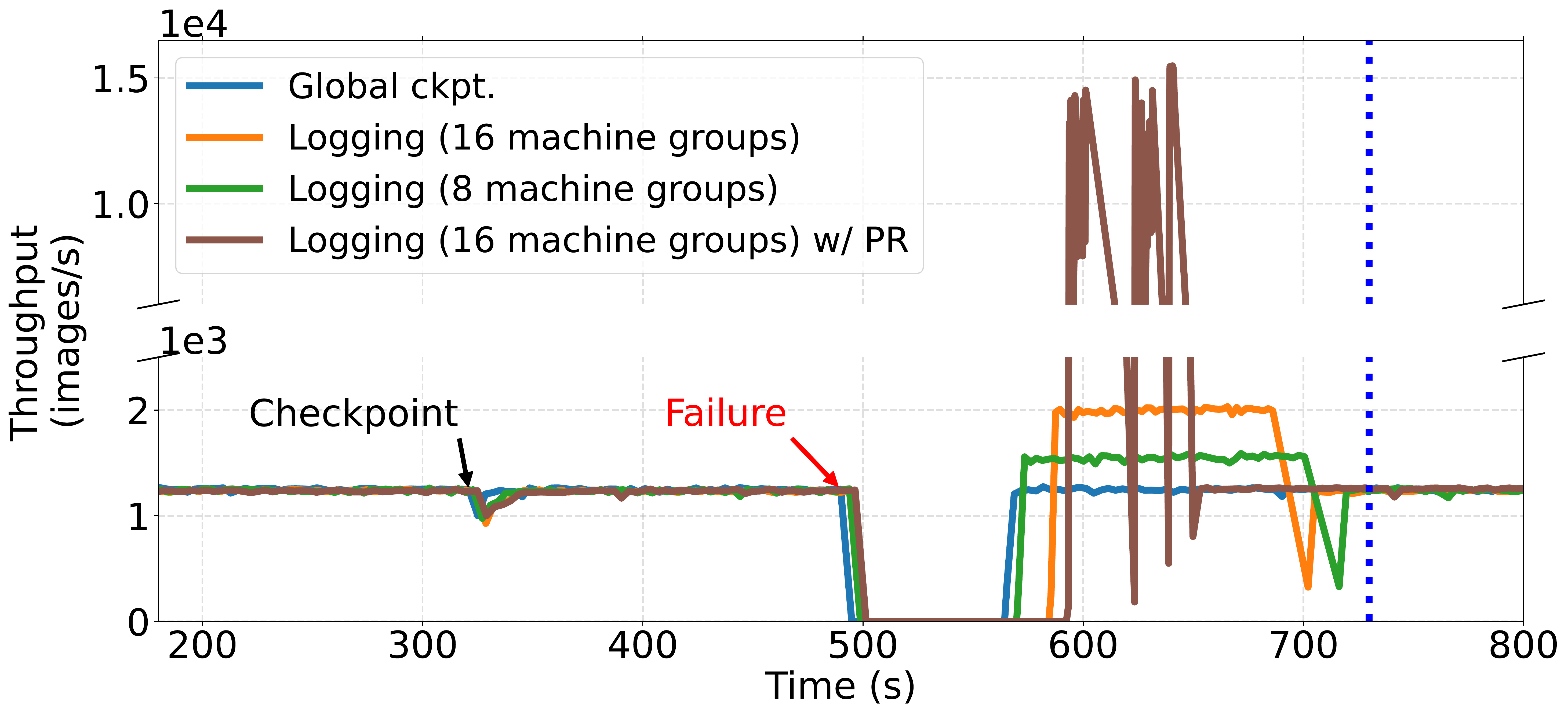}
    \caption{Training throughput of ViT-128/32 during failure recovery. Blue dashed line indicates completion of recovery with global checkpointing.}
    \label{fig:throughput_vit}
\end{figure}

\begin{table}[t]
    \centering
    \caption{Space overhead caused by logging per iteration.} 
    \begin{tabular}{cccc}
        \toprule
        \multirow{2}{*}{\footnotesize{Model}} &  \footnotesize{\#Machine} & \footnotesize{Total logging} & \footnotesize{Average consumed} \\
         &  \footnotesize{group} & \footnotesize{size (GB)} & \footnotesize{bandwidth (GB/s)}  \\
        \midrule
        \multirow{2}{*}{\footnotesize{ViT-128/32}} &\footnotesize{16}
        & \footnotesize{24.66} & \footnotesize{0.23} \\
         &\footnotesize{8}
        & \footnotesize{11.51} & \footnotesize{0.11} \\
        \multirow{2}{*}{\footnotesize{BERT-128}} &\footnotesize{16}
        & \footnotesize{8.05} & \footnotesize{0.075}  \\
         &\footnotesize{8}
        & \footnotesize{3.76} & \footnotesize{0.035}  \\
        \bottomrule
    \end{tabular}
    \label{tab:logging-size}
\end{table}

\vspace{1mm}
\noindent\textbf{Machine group size.} Figures \ref{fig:micro_benchmark_speedup_vit} and \ref{fig:micro_benchmark_speedup_bert} also show the impact of different machine group sizes on training throughput and recovery time for logging-based recovery. In Figure~\ref{fig:micro_benchmark_speedup_vit}, with 8 machine groups, the throughput is similar to global checkpointing, due to less logging data than with 16 machine groups (\S\ref{sec:4.3:selective_logging}). In Figure~\ref{fig:throughput_vit} and Figure~\ref{fig:micro_benchmark_speedup_vit}, we observe that logging with 8 machine groups requires a longer recovery time due to recovering a 16-stage sub-pipeline on two machines instead of the 8-stage sub-pipeline in the case of 16 machine groups. Table~\ref{tab:logging-size} shows the total logging size per iteration and average bandwidth taken by logging in bubble time with different models and different numbers of machine groups. This shows the trade-off between recovery time and space overhead with selective logging (\S\ref{sec:4.3:selective_logging}). 

\vspace{1mm}
\noindent\textbf{Parallel Recovery.} For logging with parallel recovery (\S\ref{sec:4.2:parallel_recovery}) cases in Figures~\ref{fig:throughput_vit}, \ref{fig:micro_benchmark_speedup_vit}, and \ref{fig:micro_benchmark_speedup_bert}, we use 16 workers (GPUs) to concurrently do the recovery computation for one failed worker. We see that parallel recovery significantly improves training throughput for ViT-128/32 from 12.5x to 15x as compared to global checkpointing (similar results for BERT-128) due to reduced recovery time (by 57.3\% and 76.3\% for  ViT-128/32 and BERT-128, respectively). The throughput fluctuation with parallel recovery is because parallel recovery is so fast that file transfer becomes a bottleneck, i.e., the new logging files are not yet downloaded from HDFS while the replay is already done with the earlier files. 

\begin{figure}[t]
    \centering
    \includegraphics[width=\columnwidth]{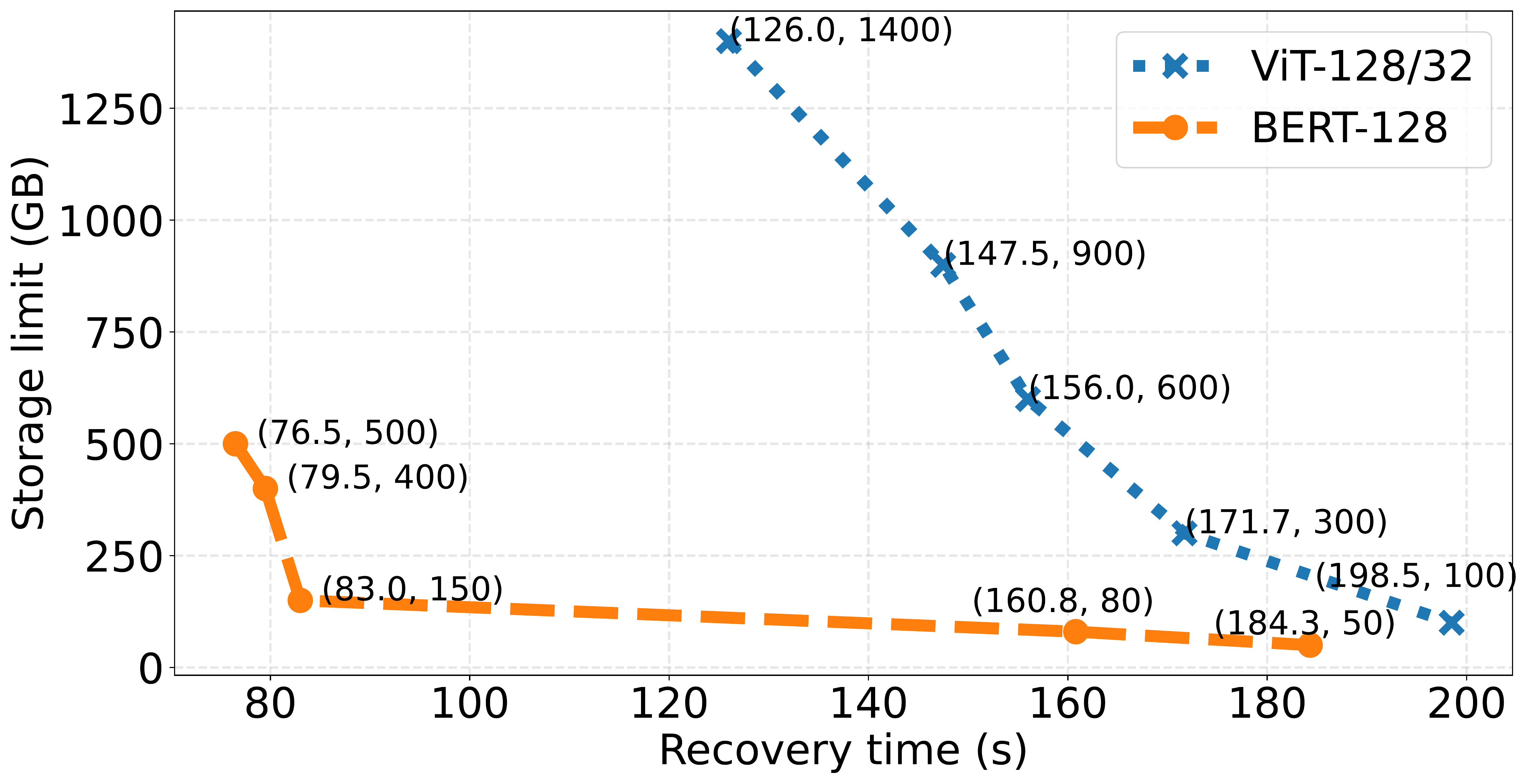}
    \caption{Trade-off between recovery time and 
    storage space limit. 
    Marker: (recovery time in seconds, storage limit in gigabytes).} 
    \label{fig:pareto}
\end{figure}

\vspace{1mm}
\noindent\textbf{Space-time trade-off.} Figure~\ref{fig:pareto} further evaluates the trade-off between recovery time and space overhead with selective logging. Given a maximal storage capacity, we use the algorithm in \S\ref{sec:4.3:selective_logging} to decide how to group machines. For both DNN models, the recovery time becomes longer when we lower the space threshold. We can identify good trade-offs using the plotted curves in practical usage. The grouping configurations can be found in Appendix~\ref{sec:grouping_details}.

\subsection{End-to-end Training} \label{sec:e2e}
\begin{figure}[!t]
   \centering
     \begin{subfigure}[b]{0.48\columnwidth}
        \centering
        \includegraphics[width=\columnwidth]{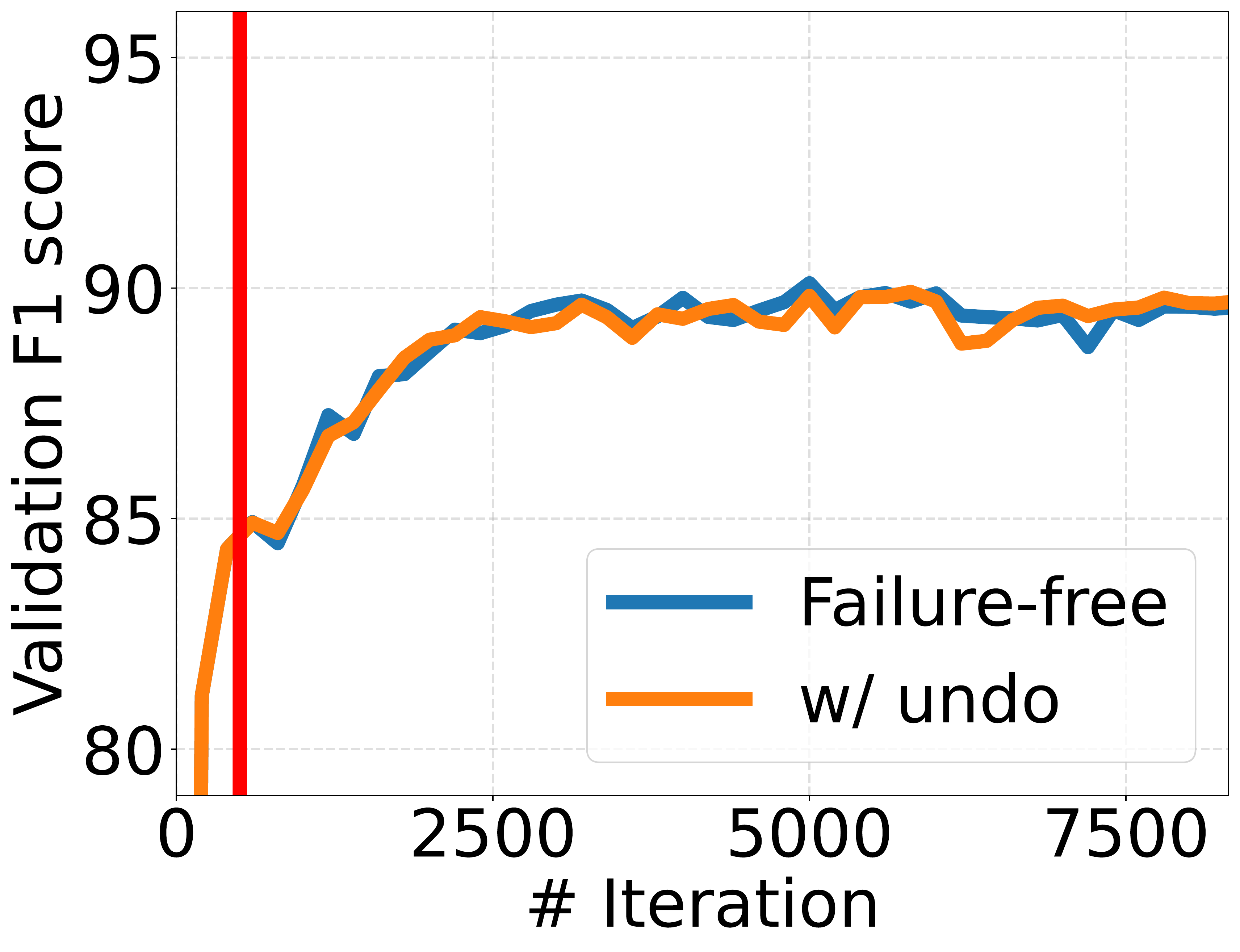}
        \caption{BERT-Large}
        \label{fig:e2e_bert}
     \end{subfigure}
     \hfill
     \begin{subfigure}[b]{0.48\columnwidth}
        \centering
        \includegraphics[width=\columnwidth]{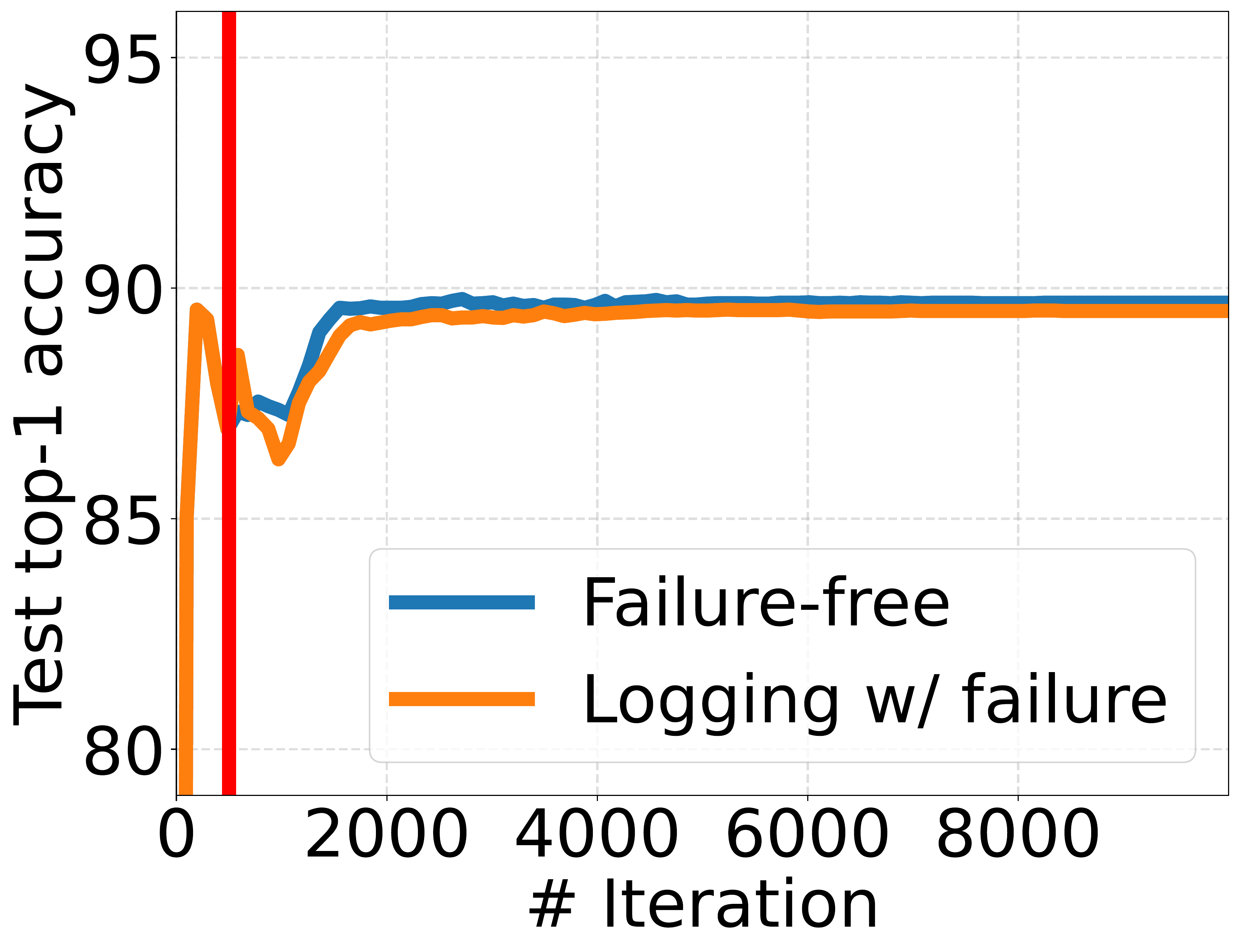}
        \caption{ViT-Base/32}
        \label{fig:e2e_vit}
     \end{subfigure}
     \caption{End-to-end training. The red line indicates a failure at iteration 500.}
     \label{fig:e2e}
\end{figure}

We next run end-to-end training to verify that \sysname{} does not affect the trained model accuracy. In Figure~\ref{fig:e2e_bert}, We finetune BERT-Large~\cite{devlin2018bert} with the Adam optimizer on SQuAD-v1.1 dataset~\cite{rajpurkar2016squad}, using pipeline parallelism with 8 GPUs on two machines. We disable logging in this experiment but inspect potential impact of update-undo (\S\ref{sec:undo_update}). We kill one machine at the end of iteration 500, intentionally make an additional update at iteration 500 and then undo this update. We observe that update-undo does not affect the final finetuning accuracy. 

In Figure~\ref{fig:e2e_vit}, we finetune ViT-Base/32~\cite{dosovitskiy2020image} using SGD with momentum on CIFAR-100 dataset for 10000 iterations, using pipeline parallelism with 12 GPUs on three machines. We kill the machine hosting stages in the middle of the pipeline (i.e., machine 1 with workers from rank 4 to rank 7) at the end of iteration 500. We do not group the machines for logging nor enable parallel recovery. We see that our logging-based failure recovery has no loss of accuracy compared to the failure-free counterpart. 

\subsection{Simulation Study}\label{sec:simulation_study}

\begin{table}[t]
    \centering
    \caption{Training workload in the simulation study.}
    \begin{tabular}{cccc}
        \toprule
        \footnotesize{Model} & \makecell{\footnotesize{total \# of} \\ \footnotesize{iterations}} & \makecell{\footnotesize{checkpoint} \\ \footnotesize{interval}} & \makecell{\footnotesize{
        End-to-end training} \\ \footnotesize{time w/o failure}} \\
        \midrule
        \footnotesize{Wide-ResNet-50} & \footnotesize{450,360} & \footnotesize{5,004} & \footnotesize{479.4hr} \\
        \footnotesize{ViT-128/32} & \footnotesize{93,600} & \footnotesize{312} & \footnotesize{85.6hr} \\
        \footnotesize{BERT-128} & \footnotesize{500,000} & \footnotesize{5,000} & \footnotesize{461.1hr} \\
        \bottomrule
    \end{tabular}
    \label{tab:simulation_info}
\end{table}

\begin{table}[t]
    \centering
    \caption{Simulated end-to-end training time with failures.}
    \begin{tabular}{ccccc}
        \toprule
        \footnotesize{Model} & \footnotesize{\#failure} & \footnotesize{Global ckpt.} & \footnotesize{\sysname{}} & \footnotesize{Speedup} \\
        \midrule
        \footnotesize{Wide-ResNet-50} & \footnotesize{28} & \footnotesize{557.4hr} & \footnotesize{480.7hr} & 1.16x \\
        \footnotesize{ViT-128/32} & \footnotesize{5} & \footnotesize{86.4hr} & \footnotesize{86.0hr} & 1.01x \\
        \footnotesize{BERT-128} & \footnotesize{27} & \footnotesize{524.2hr} & \footnotesize{476.1hr} & 1.10x \\
        \bottomrule
    \end{tabular}
    \label{tab:simulation_results}
\end{table}

We further investigate the effects of \sysname{} on the end-to-end training time through simulations. Simulation settings are given in Table~\ref{tab:simulation_info} (others are the same as experimental settings in \S\ref{sec:micro_benchmarks}). We calculate the expected end-to-end training time without failures based on the iteration time measured in the experiments and the total number of training iterations. For Wide-ResNet-50 and ViT-128/32, we assume storing a checkpoint at the end of each epoch, following common ML practice~\cite{he2016identity, zagoruyko2016wide, dosovitskiy2020image}. For BERT-128, we assume performing checkpointing once every 5000 iterations, which is 1\% of its total number of training iterations. We then inject failures uniformly randomly during training, assuming a 17-hour median-time-between-failure  (following~\cite{maeng2021understanding}). We repeat each simulation ten times and present the average results.

\vspace{1mm}
\noindent\textbf{End-to-end training time.} As shown in Table~\ref{tab:simulation_results}, \sysname{} can reduce the end-to-end training time significantly for long-running jobs, as compared to global checkpointing. Specifically, \sysname{} can speed up end-to-end training for training Wide-ResNet-50 on ImageNet and pretraining BERT-128 on the Wikipedia dataset by 1.16x and 1.10x, respectively. This translates into saving 77 hours and 48 hours of training time. Short-running jobs like training ViT-128/32 on ImageNet encounter fewer failures and thus benefit less from fast failure recovery. We also compare the end-to-end training time of Wide-ResNet-50 with Elastic Horovod and CheckFreq. We consider the overhead of snapshots in our simulations using data collected in \S\ref{sec:micro_benchmarks} and use the same snapshot frequency as in \S\ref{sec:micro_benchmarks}. End-to-end training with CheckFreq takes 518.9 hours, and with Elastic Horovod takes 515.9 hours. \sysname{} is 1.08 and 1.07 times faster than CheckFreq and Elastic Horovod, respectively.  

\begin{figure}[t]
    \centering
    \begin{subfigure}[b]{0.48\columnwidth}
        \centering
        \includegraphics[width=\columnwidth]{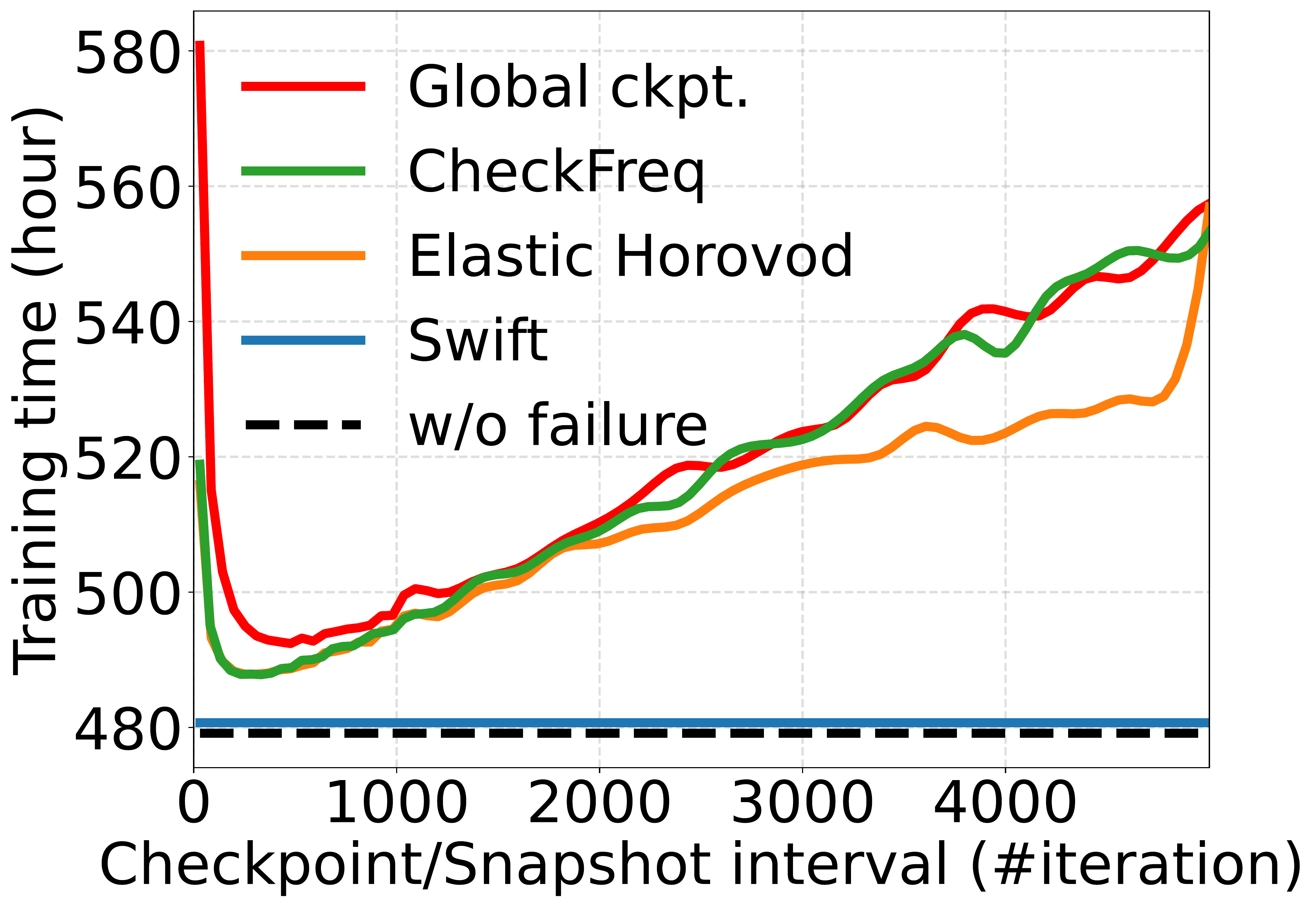}
        \caption{Wide-ResNet-50}
        \label{fig:simulation_checkpoint_frequency_resnet}
    \end{subfigure}
    \hfill
    \begin{subfigure}[b]{0.48\columnwidth}
        \centering
        \includegraphics[width=\columnwidth]{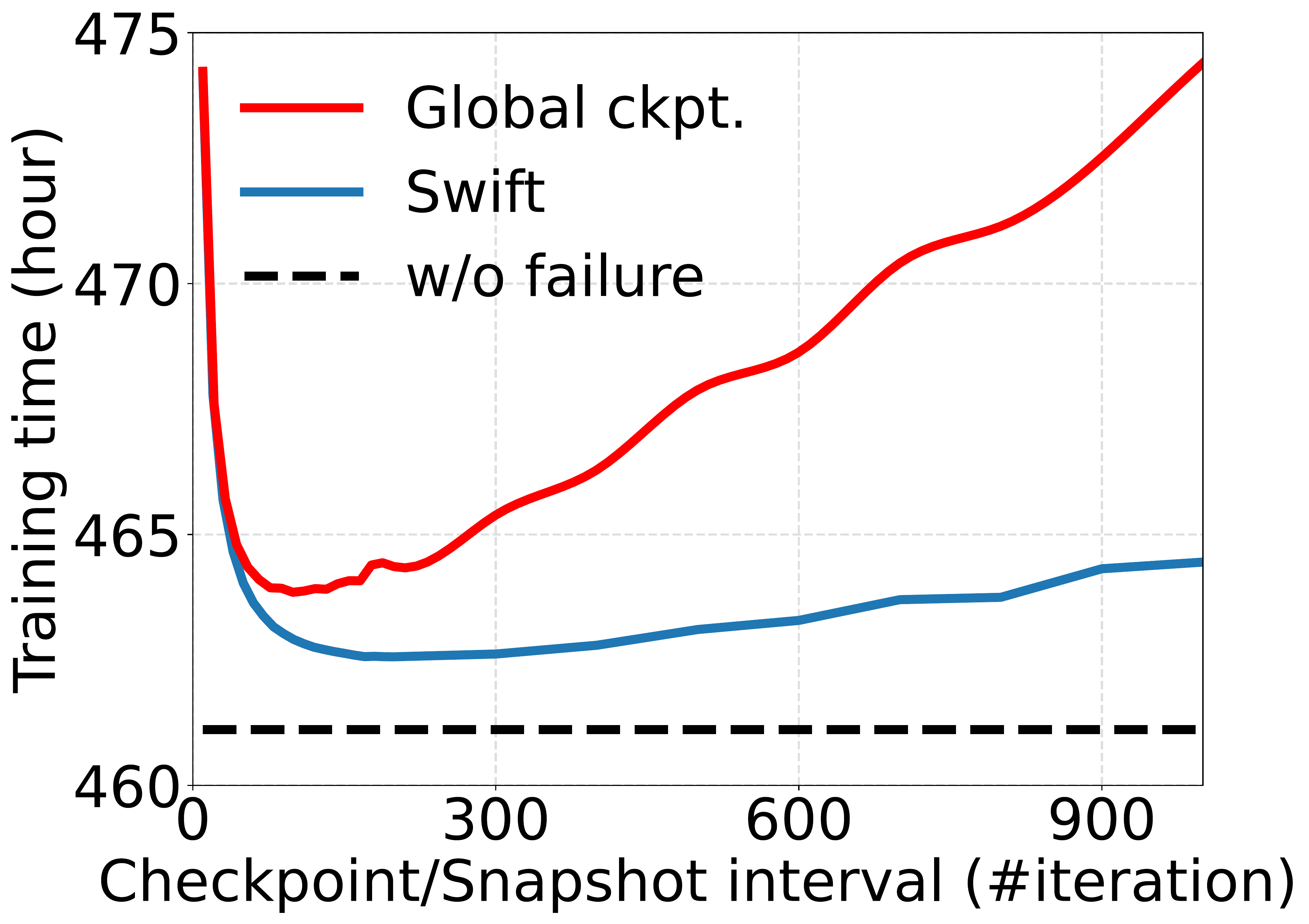}
        \caption{BERT-128}
        \label{fig:simulation_checkpoint_frequency_bert}
    \end{subfigure}
    \caption{Impact of checkpoint frequency.}
    \label{fig:simulation_checkpoint_frequency}
\end{figure}
\begin{figure}[t]
    \centering
    \begin{subfigure}[b]{0.48\columnwidth}
        \centering
        \includegraphics[width=\columnwidth]{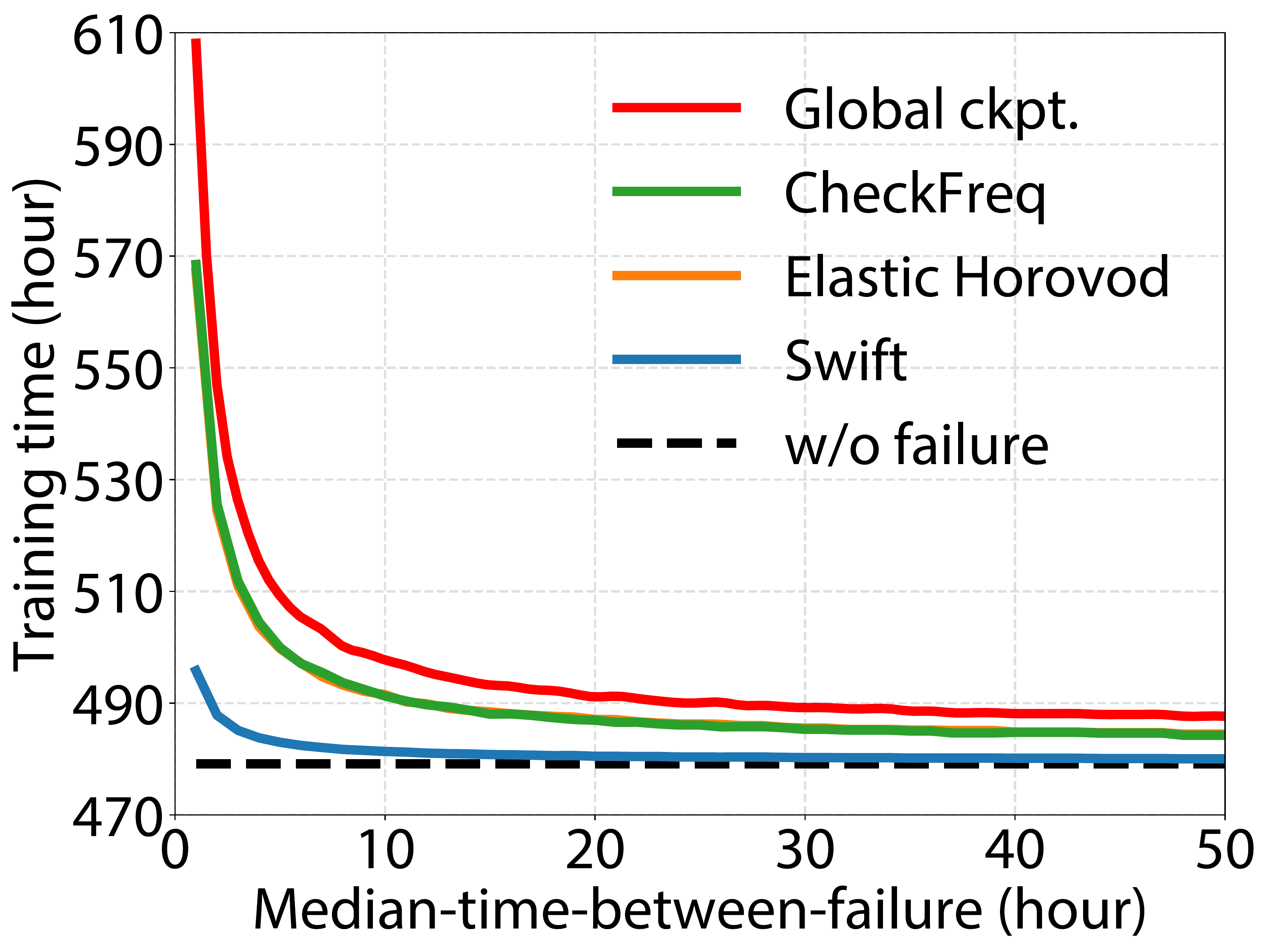}
        \caption{Wide-ResNet-50}
        \label{fig:simulation_failure_frequency_resnet}
    \end{subfigure}
    \hfill
    \begin{subfigure}[b]{0.48\columnwidth}
        \centering
        \includegraphics[width=\columnwidth]{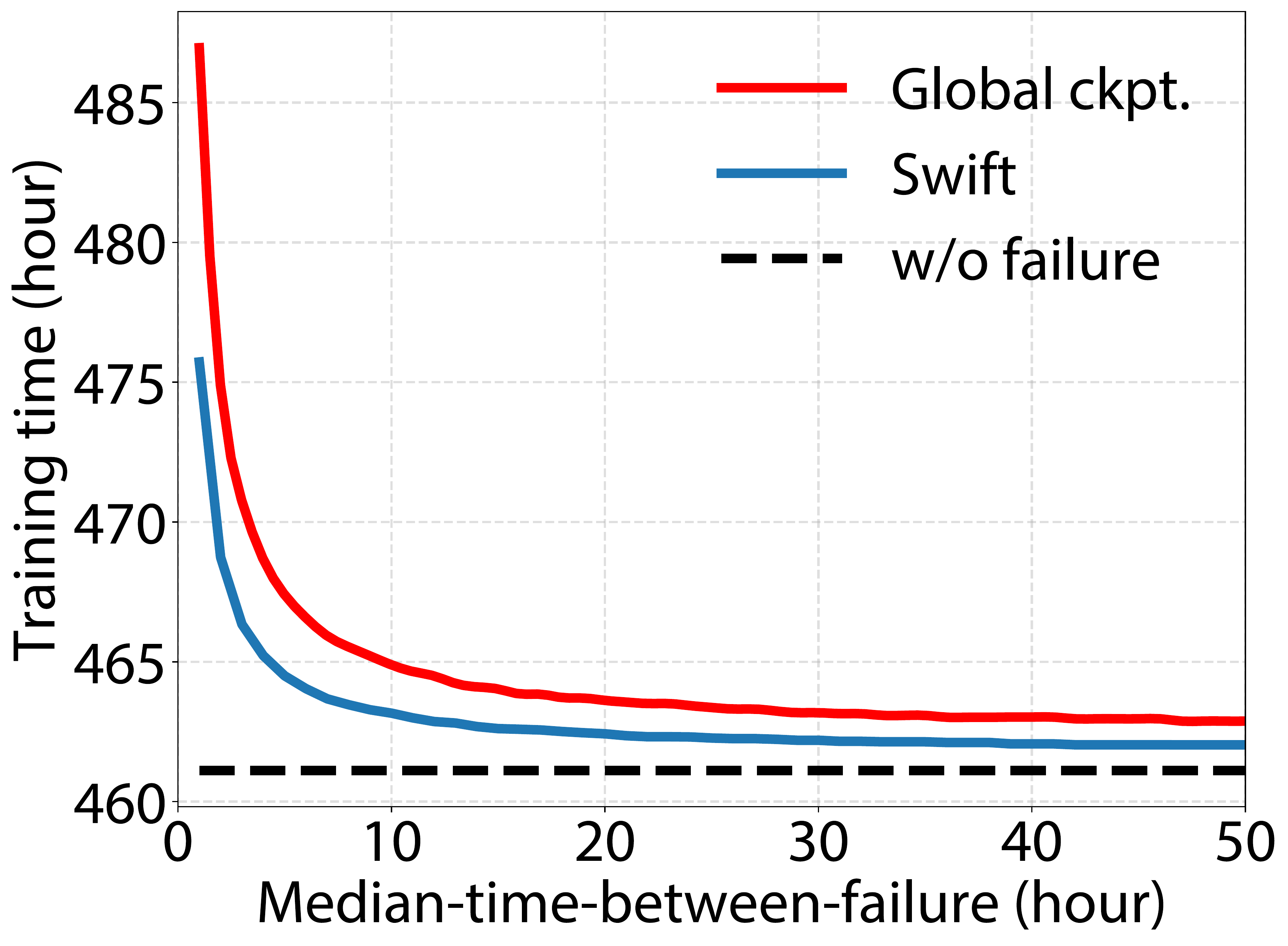}
        \caption{BERT-128}
        \label{fig:simulation_failure_frequency_bert}
    \end{subfigure}
    \caption{Impact of failure frequency.}
    \label{fig:simulation_failure_frequency}
\end{figure}

\vspace{1mm}
\noindent\textbf{Effects of checkpoint frequency.} We vary the checkpoint/snapshot frequency to investigate its impact on end-to-end training time. We keep the checkpoint frequency unchanged for replication-based recovery in \sysname{} since it does not require frequent checkpointing. As shown in Figure~\ref{fig:simulation_checkpoint_frequency}, \sysname{} achieves a shorter training time than other methods in all cases. An optimal checkpoint frequency can be obtained for each method from the curves, which leads to the shortest training time. Comparing the optimal cases of each method, for Wide-ResNet-50, \sysname{} saves 11.8 hours, 7.1 hours, and 7.2 hours compared to global checkpointing, CheckFreq, and Elastic Horovod, respectively; for BERT-128, \sysname{} saves 1.3 hours compared to global checkpointing - limited improvement due to the minimal checkpointing overhead of BERT-128 (0.93 seconds).

\vspace{1mm}
\noindent\textbf{Effects of failure frequency.} We further adjust the median-time-between-failure to investigate its effect on end-to-end training time while fixing the checkpoint/snapshot frequency to the optimal frequencies given by Figure~\ref{fig:simulation_checkpoint_frequency}. Figure~\ref{fig:simulation_failure_frequency} shows that \sysname{} achieves better speedup when failures are more frequent and also the shortest training time among all methods when failures are infrequent.

%% file: related_work.tex
\section{Related Work}
\vspace{1mm}
\noindent\textbf{Elastic training.} Most DL jobs use static job configuration (e.g., the number of workers). In elastic training, workers can join and leave. The job can scale out to utilize transient idle resources (e.g., spot instances in cloud computing), or scale in to reserve resources for high-priority jobs. Unfortunately, most elastic training works~\cite{mai2020kungfu, peng2018optimus, qiao2021pollux} still rely on checkpoint-restart method to avoid the crash-consistency problem (\S\ref{sec2.3:crash_consistency problem}). \sysname{} can resolve the inconsistency using update-undo (\S\ref{sec:undo_update}) and thus benefit elastic training (e.g., broadcast the worker's state when new workers come in).

\vspace{1mm}
\noindent\textbf{Checkpointing in DL systems.} Check-N-Run~\cite{eisenman2022check} proposes incremental checkpointing tailored for training DL recommendation models, exploiting the fact that only a fraction of the recommendation model is updated in each iteration. It is complementary to our work because \sysname{} is not limited to recommendation models. The MLP layers in recommendation models are usually trained using data parallelism~\cite{maeng2021understanding}, which can benefit from the replication-based recovery. Orpheus~\cite{xie2018orpheus} also proposes incremental checkpointing but stores sufficient vectors of gradients, which are much smaller than the gradients themselves. During recovery, the gradients are reconstructed by the stored sufficient vectors and applied to a checkpoint to recover the model state. Our logging method can be seen as an extension of their approach. We consider the sufficient vectors (e.g., intermediate activation/gradient in pipeline parallelism) of the computation graph on a machine rather than a single operator. We log data asynchronously by upstream machines while they require costly synchronous logging for consistency. Recent works also propose partial recovery that loads the checkpoint of the failed machine only and continues with training~\cite{qiao2019fault, maeng2021understanding}. Partial recovery avoids global rollback but incurs accuracy loss due to inconsistent model state among workers~\cite{maeng2021understanding}. In contrast, \sysname{} does not degrade final model accuracy while reducing recovery time.

\vspace{1mm}
\noindent\textbf{Large-scale DNN training.} In addition to parallelism, other complementary techniques for large-scale DNN training include memory optimization~\cite{zero-infinity} and mixed-precision training~\cite{micikevicius2018mixed}. \sysname{} can be combined with many of them. For example, we can combine our replication-based recovery with Fully Sharded Data Parallel (FSDP), a popular memory optimization technique that shards the model state across data-parallel workers~\cite{zero-infinity}. We can maintain two copies of each piece of the sharded model state for failure resilience. Moreover, mixed-precision training can reduce the logging size due to using a lower precision for intermediate data~\cite{micikevicius2018mixed}. 

%% file: appendix.tex
\appendices
\section{Update-undo Algorithms}
\label{sec:appendix_a}

\begin{algorithm}[H]
    \caption{SGD}
    \label{alg:sgd}
    \begin{algorithmic}[1]
\State \textbf{Input:} learning rate sequence $\{\eta_t\}^{T}_{t=1}$; weight decay $\lambda > 0$.
\State \textbf{Initialize:} $x_1 \in \mathbb{R}^d$.

\State \textbf{for} $i=1$ \textbf{to} $T$ \textbf{do}
    \State \hspace*{\algorithmicindent} $g_t = \nabla f(x_t)$
    \State \hspace*{\algorithmicindent} $x_{t+1} = x_t - \eta_t (g_t + \lambda x_t)$
\State \textbf{end for}

    \end{algorithmic}
\end{algorithm}

\begin{algorithm}[H]
    \caption{Undo SGD}
    \label{alg:undo-sgd}
    \begin{algorithmic}[1]
\State \textbf{Input:} learning rate sequence $\{\eta_t\}^{T}_{t=1}$; weight decay $\lambda > 0$; $x_{t+1} \in \mathbb{R}^d$; $g_t \in \mathbb{R}^d$.
    \State  $x_t = (x_{t+1} + \eta_t g_t) / (1 - \eta \lambda)$
    \end{algorithmic}
\end{algorithm}







\begin{algorithm}[H]
    \caption{Adam~\cite{kingma2014adam}}
    \label{alg:adam}
    \begin{algorithmic}[1]
\State \textbf{Input:} learning rate sequence $\{\eta_t\}^{T}_{t=1}$; weight decay $\lambda > 0$; Exponential decay rates for moment estimates \ $0\leq \beta_1, \beta_2 \textless 1$; $\epsilon=10^{-8}$.
\State \textbf{Initialize:} $x_1 \in \mathbb{R}^d$; $m_0 = 0 \in \mathbb{R}^d$; $v_0 = 0\in \mathbb{R}^d$.

\State \textbf{for} $i=1$ \textbf{to} $T$ \textbf{do}
    \State \hspace*{\algorithmicindent}$g_t = \nabla f(x_{t})$
    \State \hspace*{\algorithmicindent}$g_t^{\prime} = g_t + \lambda \cdot x_t$
    \State \hspace*{\algorithmicindent}$m_t = \beta_1 \cdot m_{t-1} + (1 - \beta_1) \cdot g_t^{\prime}$
    \State \hspace*{\algorithmicindent}$v_t = \beta_2 \cdot v_{t-1} + (1 - \beta_2) \cdot g_t^{{\prime}2}$
    \State \hspace*{\algorithmicindent}$\widehat{m}_t = m_t / (1 - \beta_1^t)$
    \State \hspace*{\algorithmicindent}$\widehat{v}_t = v_t / (1 - \beta_2^t)$
    \State \hspace*{\algorithmicindent}$x_t = x_{t-1} - \eta_t \cdot \widehat{m}_t / (\sqrt{\widehat{v}_t} + \epsilon)$
\State \textbf{end for}

    \end{algorithmic}
\end{algorithm}

\begin{algorithm}[H]
    \caption{Undo Adam}
    \label{alg:undo-adam}
    \begin{algorithmic}[1]
\State \textbf{Input:} learning rate sequence $\{\eta_t\}^{T}_{t=1}$; weight decay $\lambda > 0$; Exponential decay rates for moment estimates \ $0\leq \beta_1, \beta_2 \textless 1$; $\epsilon=10^{-8}$; $x_{t+1} \in \mathbb{R}^d$; $g_t \in \mathbb{R}^d$; $m_t \in \mathbb{R}^d$; $v_t \in \mathbb{R}^d$.

\State $\widehat{m}_t = m_t / (1 - \beta_1^t)$
\State $\widehat{v}_t = v_t / (1 - \beta_2^t)$
\State $x_t = x_{t+1} + \eta_t \cdot \widehat{m}_t / (\sqrt{\widehat{v}_t} + \epsilon)$
\State $g_t^{\prime} = g_t + \lambda \cdot x_t$
\State $m_{t-1} = (m_t - (1 - \beta_1) \cdot g_t^{\prime}) / \beta_1$
\State $v_{t-1} = (v_t - (1 - \beta_2) \cdot g_t^{\prime}2) / \beta_2$
    \end{algorithmic}
\end{algorithm}

\begin{algorithm}[H]
    \caption{AdamW~\cite{loshchilov2017decoupled}}
    \label{alg:adamw}
    \begin{algorithmic}[1]
\State \textbf{Input:} learning rate sequence $\{\eta_t\}^{T}_{t=1}$; weight decay $\lambda > 0$; Exponential decay rates for moment estimates \ $0\leq \beta_1, \beta_2 \textless 1$; $\epsilon=10^{-8}$.
\State \textbf{Initialize:} $x_1 \in \mathbb{R}^d$; $m_0 = 0 \in \mathbb{R}^d$; $v_0 = 0\in \mathbb{R}^d$.

\State \textbf{for} $i=1$ \textbf{to} $T$ \textbf{do}
    \State \hspace*{\algorithmicindent}$g_t = \nabla f(x_{t})$
    \State \hspace*{\algorithmicindent}$m_t = \beta_1 \cdot m_{t-1} + (1 - \beta_1) \cdot g_t$
    \State \hspace*{\algorithmicindent}$v_t = \beta_2 \cdot v_{t-1} + (1 - \beta_2) \cdot g_t^2$
    \State \hspace*{\algorithmicindent}$\widehat{m}_t = m_t / (1 - \beta_1^t)$
    \State \hspace*{\algorithmicindent}$\widehat{v}_t = v_t / (1 - \beta_2^t)$
    \State \hspace*{\algorithmicindent}$x_t = x_{t-1} - \eta_t \cdot (\alpha \cdot \widehat{m}_t / (\sqrt{\widehat{v}_t} + \epsilon) + \lambda \cdot x_{t-1})$
\State \textbf{end for}

    \end{algorithmic}
\end{algorithm}

\begin{algorithm}[H]
    \caption{Undo AdamW}
    \label{alg:undo-adamw}
    \begin{algorithmic}[1]
\State \textbf{Input:} learning rate sequence $\{\eta_t\}^{T}_{t=1}$; weight decay $\lambda > 0$; Exponential decay rates for moment estimates \ $0\leq \beta_1, \beta_2 \textless 1$; $\epsilon=10^{-8}$; $x_1 \in \mathbb{R}^d$; $m_0 = 0 \in \mathbb{R}^d$; $v_0 = 0\in \mathbb{R}^d$.

    \State $\widehat{m}_t = m_t / (1 - \beta_1^t)$
    \State $\widehat{v}_t = v_t / (1 - \beta_2^t)$
    \State $x_{t-1} = (x_t + \eta_t \cdot (\alpha \cdot \widehat{m}_t / \sqrt{\widehat{v}_t} + \epsilon)) / (1 - \eta_t \cdot \lambda)$
    \State $m_{t-1} = (m_t - (1 - \beta_1) \cdot g_t) / \beta_1$
    \State $v_{t-1} = (v_t - (1 - \beta_2) \cdot g_t^2) / \beta_2$
    
    \end{algorithmic}
\end{algorithm}

\section{Support multiple failures and cascading failures} 
\label{sec:multiple_cascading_failures}
\sysname{} can tolerate multiple simultaneous failures. Even if there are multiple failures, replication-based recovery can be used if there is still a copy of the lost model state. For logging-based recovery, if the failed workers (on failure machines) constitute a consecutive portion of the pipeline, they can be recovered jointly; if they span different portions of the pipeline, we can recover those portions independently. 

\sysname{} can also tolerate cascading failures. If another failure occurs while recovering from one failure, you can use replication-based recovery as long as that lost state is still backed up. For logging-based recovery, during the recovery of a failed machine, if another machine crashes which hosts workers which are connected in the pipeline to workers on the being-recovered machine, the ongoing recovery process is aborted, and the two machines are recovered together; if the newly crashed machine does not host workers connected to those on the being-recovered machine, it does not affect the ongoing recovery process, and the new failure can be recovered independently.

\section{Grouping details}
\label{sec:grouping_details}

\begin{table*}[h]
    \centering
    \caption{BERT-128's grouping results. The number in the outcome represents the machine rank.}
    \begin{tabular}{|l|l|}
    \hline
        Storage limit (Bytes) & Outcome \\ \hline
        5.00E+11 & [[0], [1], [2], [3], [4], [5], [6], [7], [8], [9], [10], [11], [12], [13], [14], [15]] \\ \hline
        4.00E+11 & [[0], [1], [2], [3], [4], [5], [6], [7], [8], [9], [10], [11, 12], [13, 14], [15]] \\ \hline
        3.50E+11 & [[0], [1], [2], [3], [4], [5], [6], [7], [8], [9, 10], [11, 12], [13, 14], [15]] \\ \hline
        3.00E+11 & [[0], [1], [2], [3], [4], [5], [6], [7, 8], [9, 10], [11, 12], [13, 14, 15]] \\ \hline
        2.50E+11 & [[0], [1], [2], [3, 4], [5, 6], [7, 8], [9, 10], [11, 12], [13, 14, 15]] \\ \hline
        2.20E+11 & [[0], [1, 2], [3, 4], [5, 6], [7, 8], [9, 10, 11, 12], [13, 14, 15]] \\ \hline
        1.50E+11 & [[0, 1, 2], [3, 4], [5, 6, 7, 8], [9, 10, 11, 12], [13, 14, 15]] \\ \hline
        1.00E+11 & [[0, 1, 2], [3, 4, 5, 6, 7, 8], [9, 10, 11, 12, 13, 14, 15]] \\ \hline
        8.00E+10 & [[0, 1, 2, 3, 4, 5, 6, 7, 8], [9, 10, 11, 12, 13, 14, 15]] \\ \hline
        5.00E+10 & [[0, 1, 2, 3, 4, 5, 6, 7, 8, 9, 10, 11, 12, 13, 14, 15]] \\ \hline
    \end{tabular}

\end{table*}

\begin{table*}[h]
    \centering
    \caption{ViT-128/32's grouping results. The number in the outcome represents the machine rank.}
    \begin{tabular}{|l|l|}
    \hline
        Storage limit (Bytes) & Outcome \\ \hline
        1.40E+12 & [[0], [1], [2], [3], [4], [5], [6], [7], [8], [9], [10], [11], [12], [13], [14], [15]] \\ \hline
        1.20E+12 & [[0], [1], [2], [3], [4], [5], [6], [7], [8], [9], [10], [11, 12], [13, 14], [15]] \\ \hline
        1.10E+12 & [[0], [1], [2], [3], [4], [5], [6], [7], [8], [9], [10], [11, 12], [13, 14, 15]] \\ \hline
        1.00E+12 & [[0], [1], [2], [3], [4], [5], [6], [7, 8], [9, 10], [11, 12], [13, 14, 15]] \\ \hline
        9.00E+11 & [[0], [1], [2], [3], [4], [5], [6, 7], [8, 9], [10, 11, 12], [13, 14, 15]] \\ \hline
        8.00E+11 &  [[0], [1], [2], [3], [4, 5], [6, 7], [8, 9], [10, 11, 12], [13, 14, 15]] \\ \hline
        7.00E+11 &  [[0], [1], [2], [3], [4, 5], [6, 7], [8, 9], [10, 11, 12, 13, 14, 15]] \\ \hline
        6.00E+11 & [[0], [1], [2, 3], [4, 5], [6, 7], [8, 9], [10, 11, 12, 13, 14, 15]] \\ \hline
        5.00E+11 & [[0, 1], [2, 3], [4, 5], [6, 7, 8, 9], [10, 11, 12, 13, 14, 15]] \\ \hline
        4.00E+11 & [[0, 1], [2, 3, 4, 5], [6, 7, 8, 9], [10, 11, 12, 13, 14, 15]] \\ \hline
        3.00E+11 & [[0, 1], [2, 3, 4, 5], [6, 7, 8, 9, 10, 11, 12, 13, 14, 15]] \\ \hline
        2.00E+11 & [[0, 1, 2, 3, 4, 5], [6, 7, 8, 9, 10, 11, 12, 13, 14, 15]] \\ \hline
        1.00E+11 & [[0, 1, 2, 3, 4, 5, 6, 7, 8, 9, 10, 11, 12, 13, 14, 15]] \\ \hline
    \end{tabular}

\end{table*}

We use the model and settings as described in \S\ref{sec:micro_benchmarks}. We profile the computation time for 5 iterations of each stage and take the average time. And the communication size between each stage can be calculated directly based on the model configuration and batch size.